\documentclass{article}

\usepackage{graphicx}
\usepackage{wasysym}

\author{Aron C. Wall\footnote{aronwall@umd.edu}
\\ \textit{Maryland Center for Fundamental Physics} \\ \textit{Department of Physics} \\ \textit{University of Maryland} \\ \textit{College Park, MD 20740-4111, USA} }
\title{The Generalized Second Law implies a Quantum Singularity Theorem}
\date{\today}

\begin{document}

\maketitle

\begin{abstract}

The generalized second law can be used to prove a singularity theorem, by generalizing the notion of a trapped surface to quantum situations.  Like Penrose's original singularity theorem, it implies that spacetime is null geodesically incomplete inside black holes, and to the past of spatially infinite Friedmann--Robertson--Walker cosmologies.  If space is finite instead, the generalized second law requires that there only be a finite amount of entropy producing processes in the past, unless there is a reversal of the arrow of time.  In asymptotically flat spacetime, the generalized second law also rules out traversable wormholes, negative masses, and other forms of faster-than-light travel between asymptotic regions, as well as closed timelike curves.  Furthermore it is impossible to form baby universes which eventually become independent of the mother universe, or to restart inflation.  Since the semiclassical approximation is used only in regions with low curvature, it is argued that the results may hold in full quantum gravity.

The introduction describes the second law and its time-reverse, in ordinary and generalized thermodynamics, using either the fine-grained or the coarse-grained entropy.  (The fine-grained version is used in all results except those relating to the arrow of time.)
\end{abstract}

\newpage

\tableofcontents

\section{Introduction}\label{intro}

It is been speculated for some time that quantum effects will remove singularities from any theory of quantum gravity \cite{SingRes}.  Singularity resolution has been attempted in both string theory \cite{stringy} and loop quantum gravity \cite{loopy}.  Implicit in many of these works is the idea that quantum gravity will permit spacetimes to evade the classical singularity theorems of general relativity, and thus permit continuation past the would-be singularity.  The classical singularity theorems all assume certain positivity conditions on the stress-energy tensor.  However, all such conditions can be violated locally in quantum field theory.  One might therefore suppose that in the highly quantum regions near a Big Bang or black hole singularity, temporary doses of negative energy might induce a bounce, avoiding the singularity \cite{bounce}.

The question thus arises whether there is a quantum mechanical generalization of any of the singularity theorems, which would make singularities inevitable even in quantum situations.  Such a singularity theorem would have to have some assumption used in place of an energy condition which is valid in quantum situations.  In this article the (fine-grained) generalized second law (GSL) of horizon thermodynamics will be proposed as a substitute.  Since the GSL is widely believed to hold as a consequence of the statistical mechanical properties of quantum gravitational degrees of freedom \cite{BHstat}, it is a good candidate for a physical law likely to hold even in a full theory of quantum gravity.

Penrose's singularity theorem \cite{HawkingEllis} applies to classical general relativity coupled to matter obeying the null energy condition
\begin{equation}\label{NEC}
T_{ab} k^a k^b \ge 0,
\end{equation}
where $k^a$ is any null vector.  It says that on any globally hyperbolic spacetime with a noncompact Cauchy surface $\Sigma$, if there is a ``trapped surface'' $T$ on $\Sigma$ such that the outward-going null surface generated by $T$ is contracting, then the spacetime cannot be null geodesically complete.  The proof uses the Raychaudhuri equation to show that the null surface generated by $T$ must have conjugate points, but this is incompatible with the spacetime continuing any further.  The assumption that $\Sigma$ is noncompact is required to prevent the lightrays from simply intersecting one another outside of $T$.  The assumption of global hyperbolicity is required because otherwise an initially noncompact universe can evolve into a compact universe as time passes (and in fact, there are non-globally hyperbolic black hole spacetimes which satisfy all other conditions of the theorem but have a Cauchy horizon instead of singularities \cite{HawkingEllis}).

Penrose's theorem can be used to show that black holes must have singularities.  By reversing the time orientation it can also show that if the universe is spatially infinite, it must have had a Big Bang singularity somewhere \cite{HawkingEllis}.  A further consequence is that there are no traversable wormholes \cite{worm}, and that it is impossible to create an inflationary region in a laboratory without any initial singularities \cite{obstacle}.  Analogous results show that the null energy condition precludes time machines \cite{noctc} and superluminal communication \cite{olum, supercensor, woolgar94, GW00}, and requires that all asymptotically flat or AdS spacetimes to have positive ADM mass \cite{SPW, woolgar94, GW00}.

However, none of these results apply to quantum mechanical systems because all such systems violate the null energy condition (\ref{NEC}) \cite{neg}.  There are also otherwise reasonable classical theories that violate the null energy condition \cite{visser}.  Since negative mass objects probably imply that the vacuum is unstable, and time machines (and probably also wormholes) would spell trouble for causality \cite{noctc}, there ought to be some physical principle in the theory which prevents them from occurring.  This principle, unlike the null energy condition, would have to be true in quantum mechanical situations---ideally, in some complete theory of quantum gravity.

As a step in this direction, Graham and Olum \cite{GO} pointed out that the self-consistent semiclassical averaged null energy condition on achronal\footnote{A set is achronal if no two points are connected by timelike curves.} null geodesics was sufficient to rule out time machines, traversable wormholes, and negative energies.\footnote{They could not prove any singularity theorems, because these typically require the averaged null energy condition to hold on a semi-infinite null ray with one endpoint, a condition which can be violated by quantum fields.} Then \cite{anec} showed that to first order in $\hbar$, the GSL implies the condition of Graham and Olum.  This means that in any situation where quantum effects are weak, the no-go results described in \cite{GO} will follow.

In this article it will be shown more generally that the fine-grained GSL can be used to prove the inevitability of singularities, and the absence of traversable wormholes, warp drives between points at null infinity, time machines, and negative mass objects, even in quantum mechanical situations.  It will also be shown that no baby universes forming inside of black holes can be viable (in the sense of eventually becoming causally independent of the mother universe), and that it is impossible to restart inflation in the interior of an asymptotically flat (or AdS) spacetime.  

Horizon thermodynamics also severely constrains models in an FRW-like cosmology originates out of some other pre-Big-Bang universe.  In this context there are interesting constraints coming from the coarse-grained GSL, as well as the time-reverse of the fine-grained GSL. 

The various results will be shown first in the context of semiclassical gravity, in which one assumes that quantum effects are small, and can be controlled with an $\hbar$ expansion.  However, this semiclassical approximation is used only in low curvature regions far from the singularity/pathology in question.  In the high curvature region, other than the GSL itself, the results only require that basic concepts such as causality, predictivity, and topological compactness continue to have meaning in the theory of quantum gravity.  Thus it is reasonable to believe that the results will hold in a complete theory of quantum gravity.  If this conclusion is correct, we must either give up horizon thermodynamics, or else several of the dreams of science fiction fans.

Since there is no well-understood nonperturbative theory of full quantum gravity (let alone an experimentally-tested one), it is of course impossible to speak with total confidence regarding the extension of these results to this regime, which is likely of importance near singularities.  It may be that the concepts used to define the GSL apply only to semiclassical or perturbative gravity, not to the microscopic theory.  So a conservative interpretation of the restrictions is simply that such-and-such cannot occur except by means of nonperturbative quantum gravity effects.\footnote{While quantum gravity effects are expected to be important near singularities, other applications of the Penrose theorem do not require them (cf. section \ref{quantgeo}).} However, I will argue for a more expansive interpretation.

\begin{figure}[p]
\centering
\includegraphics[width=1.1\textwidth, trim=0 1.5in 0 0, clip=true]{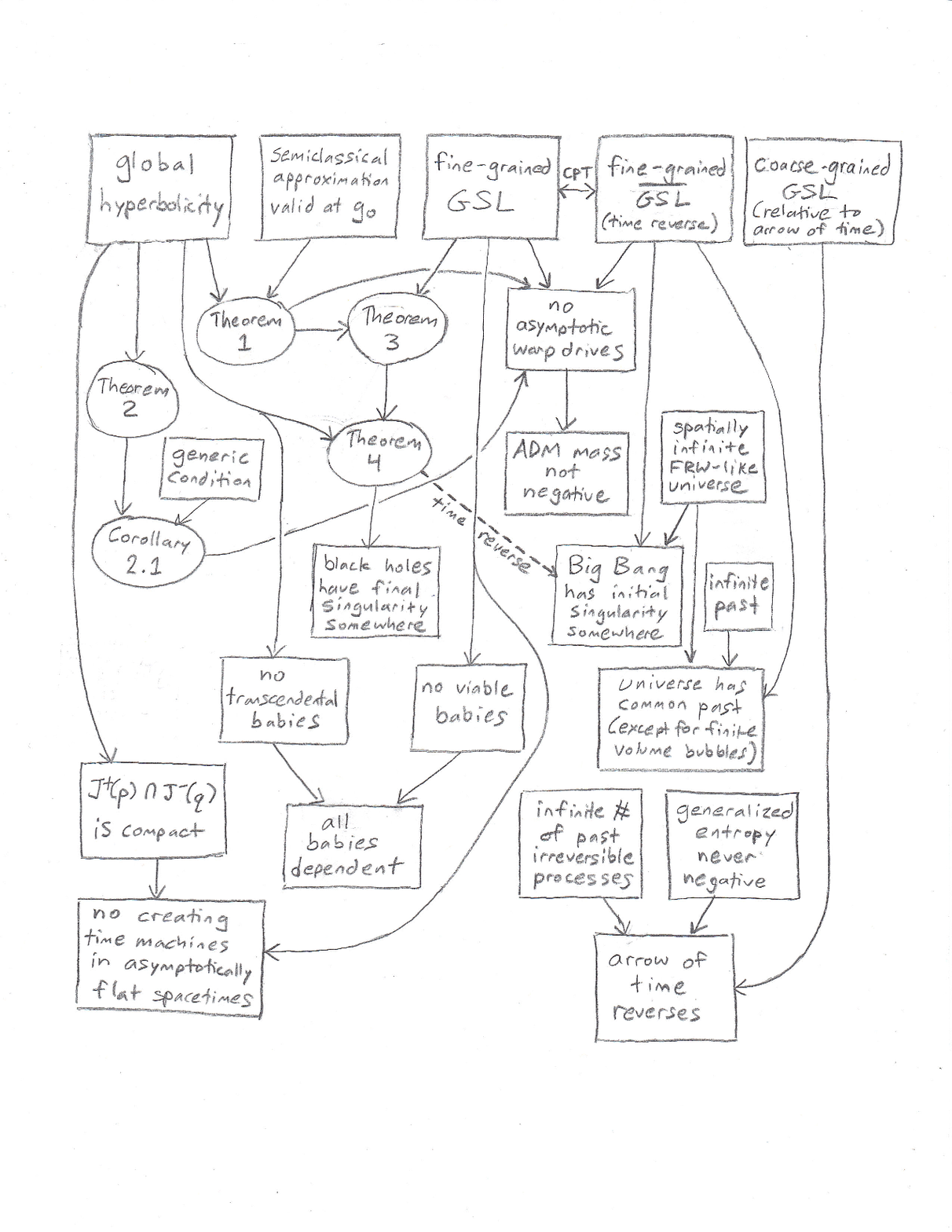}
\caption{The logical dependencies of the main hypotheses, theorems, and results in this article.  If a proposition has one or more arrows pointing to it, then the conjunction of \emph{all} propositions pointing to it is used in the proof of that proposition.  $g_0$ is a particular point in the spacetime where the semiclassical approximation must be valid; its location depends on the particular result being considered (cf. section \ref{app}).  The no-time-machines result uses the assumption that $J^+(p)\,\cap\,J^{-}(q)$ is compact for any points $p$ and $q$; since this is weaker than global hyperbolicity it is shown as following from it, although going directly from global hyperbolicity to no-time-machines is of course trivial.}\label{chart}
\end{figure}

The plan of this paper is as follows: section \ref{second} discusses the second law of thermodynamics, both in its ordinary and in its generalized form, with specific care given to the distinction between fine-grained and coarse-grained entropy, and a discussion of in which senses the second law does or does not rely on a well-defined arrow of time.  Section \ref{theorems} proves some theorems about the generalized entropy which will be used later, the most important of which is Theorem 4 which generalizes the notion of a trapped surface to quantum spacetimes.  (The casual reader may wish to skim this section).  Section \ref{app} applies the GSL to obtain the various results described in the abstract.  
Their dependency relationships are shown in Fig. \ref{chart}.

Up to this point, I will make free use of semiclassical notions of spacetime, even though such concepts are not valid in the full quantum gravity regime.  In section \ref{QG}, I will argue that the results likely apply even when quantum gravity effects are taken into consideration.  Finally, the Appendix proves a theorem used in section \ref{OSL} to help prove the ordinary second law of thermodynamics.

\section{The Second Law of Thermodynamics}\label{second}
\subsection{The Ordinary Second Law}\label{OSL}

The Ordinary Second Law (OSL) of thermodynamics states that the total entropy of a closed system cannot decrease as time passes.  For the purposes of this article I will use as my definition of the entropy of a state $\rho$ the von Neumann entropy
\begin{equation}\label{von}
S = -\mathrm{tr}(\rho\,\ln\,\rho).
\end{equation}
This is the analogue for quantum states of the classical Gibbs entropy which is defined as
\begin{equation}\label{Gibbs}
S = -\sum_i p_i\,\ln\,p_i,
\end{equation}
where $p_i$ is the probability to be in the state $i$.  Eq. (\ref{von}) reduces to Eq. (\ref{Gibbs}) whenever the density matrix $\rho$ is diagonalized.

In order to interpret the meaning of the von Neumann entropy, one needs to know whether $\rho$ is interpreted in a fine-grained sense as the complete information about a state, or in a coarse-grained sense as the information available to an observer.  In the fine-grained picture, closed quantum systems evolve by unitary evolution as time passes:
\begin{equation}\label{unitary}
\rho(t) = U(t) \rho(t_0) U^\dagger(t).
\end{equation}
Since unitary evolution does not affect the probability eigenvalues of $\rho$, this implies that
\begin{equation}
S(\rho(t)) = S(\rho(t_0)).
\end{equation}
The good news is that we have just proven the OSL because the entropy cannot be decreased.  The bad news is that although the entropy cannot decrease, it cannot increase either, so that the time-reverse of the second law, which I will denote as $\overline{\mathrm{OSL}}$, also holds.  This is because the entropy is a measure of the uncertainty in $\rho$, but the information content in $\rho$ is just the same as the uncertainty in the initial conditions.

In order to see entropy increasing, we have to move to the coarse-grained picture.  This picture can be motivated by taking a more realistic view of our ability to calculate the state at a time $t$ from the initial conditions.  If you are a Laplace's Calculator with a full understanding of the laws of nature and an infinite calculational capacity, you might well use Eq. (\ref{unitary}) to determine $\rho(t)$.  But the universe contains many complex systems for which I at any rate would be unable to specify $U$.  Given my ignorance of the exact dynamics of the universe, I cannot fully know what $\rho(t)$ is, even if I know the initial state $\rho(t_0)$.  The best I can do is rely on the things I do know about the dynamics to produce my best guess as to what I think the state is---call this $\tilde{\rho}(t)$.  Since I know that the dynamics are unitary, but I do not know the exact unitary laws of physics, I ought to be able to model my ignorance as a probability distribution over the space of possible unitary processes $U$.  This implies that I must be more uncertain about the universe at time $t$ than a Laplace's Calculator would be, so 
\begin{equation}\label{inc}
S(\tilde{\rho}(t)) \ge S(\rho(t)) = S(\rho(t_0)), \quad t > t_0.
\end{equation}
(This equation follows from the fact that entropy is a convex function, which was first proven for quantum systems by Delbr\"uck and Moli\`ere \cite{convex}.  The Appendix provides another proof following the methods of Uhlmann \cite{Uhlmann}.)

Eq. (\ref{inc}) shows that the entropy at any time must be greater than the entropy of the initial state.  This does not quite prove the OSL, because it is not yet shown whether
\begin{equation}
S(\tilde{\rho}(t_2)) \ge S(\tilde{\rho}(t_1)), \quad t_2 > t_1 > t_0.
\end{equation}
It might be, for example, that the history of the universe from $t_0$ to $t_1$ consists of some complex, calculationally intractable process, but the history of the universe from $t_1$ to $t_2$ consists of an exact reverse of that process.  Then the entropy would increase at first and then decrease again later.  In order to get the OSL, we need to know that this sort of thing does not happen in the real world, i.e. the complex processes which lead us to approximate the state of the universe with $\tilde{\rho}$ really are irreversible processes.  Another way of putting this is that once we evolve from $\rho(t_0)$ to $\tilde{\rho}(t_1)$, it must be possible to use $\tilde{\rho}(t_1)$ as a new initial condition for purposes of determining $\tilde{\rho}(t_2)$.

A trade-off has been made here.  Although the coarse-grained OSL seems to predict that the entropy will increase rather than just remain constant, by virtue of the time-reversal symmetry of the laws of physics,\footnote{or more generally, CPT symmetry.} this is only possible if there is a time-asymmetrical assumption hidden in the proof.  And there is such an assumption, embedded in the initial condition $\rho(t_0)$.  In order to get a nontrivial entropy increase, $\rho(t_0)$ must have less than the maximum possible entropy.  In other words, the universe has to have started out with low entropy.\footnote{It is also necessary not to make any restriction on the final state of the universe.  If there were a low entropy assumption made for both the initial and the final state, it would not be correct to calculate $\tilde{\rho}$ from the initial condition alone, since that would ignore additional relevant information.  This assumption is implicit in the argument for the OSL given above.}  This means that the coarse-grained OSL only holds in some states (those which really did have a low-entropy beginning), unlike the fine-grained OSL and $\overline{\mathrm{OSL}}$ which hold in every state.

The underlying time-symmetry of the argument can be illustrated by imagining that the universe had an infinite past before the ``initial'' condition.  Then for times $t_{-\infty} < t < t_0$, the same arguments given above show that entropy must be decreasing, so that the thermodynamic arrow of time is reversed.  From a thermodynamic point of view one might prefer to describe such a universe as ``beginning'' at $t_0$ and then evolving ``forwards'' in time in both directions from $t_0$.  

In order to have entropy increase for all time even with an infinite past, one might try to impose the initial condition at at $t_{-\infty}$ instead of $t_0$.  However, one would then expect that the universe would have already arrived at thermal equilibrium by any finite time $t$---assuming that there are an infinite number of potentially entropy producing processes before the time $t$.

This conclusion might be evaded if the laws of physics permit the total entropy of the universe to increase indefinitely without ever coming to equilibrium.  This might actually be the case in theories of gravity similar to general relativity \cite{carroll}.   In an expanding universe the total volume of space can grow without limit.  In classical general relativity, black holes have zero temperature, and can therefore store an arbitrarily large amount of entropy using an arbitrarily small amount of energy.  Even semiclassically, it is possible for a thermal black hole to absorb an arbitrarily large amount of entropy, if it is critically illuminated for a long period of time \cite{FPST}.\footnote{Here we are talking about the \emph{ordinary} entropy of the interior of the black hole, not the \emph{generalized} entropy of its horizon (which is the subject of the next section).  The latter is bounded at any finite energy; the former might not be, depending on ones views about entropy bounds and black hole information loss.  My claims in this article do not require taking a stand on this controversy, since they will be based on the generalized entropy rather than the ordinary entropy.}  Then in an infinitely large cosmos, one might have an entropy which is both infinite and increasing at all times.

Quantum modifications to general relativity may lead to even more exotic possibilities for cosmologies in an eternal-steady state entropy increase.  One proposal is that each universe can spawn new universes \cite{babies, FMM}, each of which might continue to increase in entropy without any violation of the OSL.  It has even been suggested that baby universes may have slightly different laws of physics leading to Darwinian adaptation of universes \cite{smolin}.  Or the universe might go through a series of cycles of de Sitter expansion and thus grow its volume and entropy without limit, as in the ekpyrotic model \cite{ek}.  Are such pictures possible?  In order to answer that question, we will now explore the generalization of thermodynamics to gravitational systems.

\subsection{The Generalized Second Law}\label{GSL}

One comparatively simple modification which must be made to the laws of thermodynamics when taking gravity into account, is that there is no longer an absolute notion of time; there are many equally good ``t'' coordinates that can be used.  Since the OSL above was formulated in terms of a ``t'' coordinate, it is necessary to modify the OSL by considering evolution from an arbitrary complete spatial slice $\Sigma$ to a complete spatial slice $\Sigma^\prime$ which is nowhere to the past of $\Sigma$.  One can then formulate the OSL as the statement that the von Neumann entropy of $\Sigma^\prime$ must be at least as great as the entropy of $\Sigma$.

But there is a more profound modification to thermodynamics which arises for quantum fields in gravitational settings, which is that the laws seem to still apply in the case of certain \emph{open} systems.  One example of such a system is the exterior of a black hole.  In this case there are gedankenexperiments \cite{gedanken} and partial proofs \cite{10proofs, myproofs} which show that the generalized entropy, defined as follows, is nondecreasing with time:
\begin{equation}
S_\mathrm{gen} = S_{\mathrm{H}} + S_\mathrm{out}.
\end{equation}
Here $S_\mathrm{out}$ is the entropy of everything outside the black hole and $S_{\mathrm{H}}$ is the entropy of the horizon itself, each defined on the same spatial slice $\Sigma$.  $S_{\mathrm{H}}$ depends on the gravitational Lagrangian \cite{noether}, and for general relativity takes the form
\begin{equation}\label{BH}
S_{\mathrm{H}} = \frac{A}{4\hbar G}.\footnote{We will refer to $S_\mathrm{H}$ as the ``horizon entropy'' even when it is evaluated on surfaces that are not horizons.} 
\end{equation}  
Na\"{i}vely one might have thought that one could make the entropy outside of a black hole go down by simply throwing entropy across the event horizon.  But such entropy tends to be accompanied by energy, which in turn increases the mass of the black hole and correspondingly $S_{\mathrm{H}}$.  Similarly, Hawking radiation reduces the size of the black hole but the decrease in $S_\mathrm{H}$ is compensated for by the increase of entropy outside of the black hole \cite{hawking}.

There are some nuances in the definition of $S_\mathrm{gen}$.  $S_\mathrm{out}$ includes a divergent contribution coming from the short-distance entanglement entropy of quantum fields near the horizon.  This divergence is quadratic with respect to a UV length cutoff.  Thus $S_\mathrm{out}$, defined as the von Neumann entropy (\ref{von}) is formally infinite and requires renormalization.  A second issue is that in perturbative quantum gravity, renormalization should lead to higher order terms in the Lagrangian which renormalize Newton's constant $G$ and also add higher order curvature terms, leading to cutoff-dependent corrections to $S_{\mathrm{H}}$.  The good news is that these two problems seem to cancel each other out---i.e. the divergence in $S_\mathrm{out}$ can be absorbed into the coupling constants that appear in $S_{\mathrm{H}}$.  This has been shown to one loop order for certain scalar and spinor theories \cite{renorm}, but there is an additional term appearing in the horizon entropy for gauge theories which is still not well understood \cite{kabat}.

Another important question is whether the GSL applies to any other horizons besides black hole event horizons.  The answer seems to be yes: horizon thermodynamics seems to apply to de Sitter and Rindler horizons as well \cite{JP, myproofs}.  However, the GSL does not hold on all null surfaces \cite{10proofs}; for example the past lightcone of a point has decreasing area classically, leading to a $\mathcal{O}(\hbar^{-1})$ decrease in the generalized entropy, but the increase in $S_\mathrm{out}$ due to quantum effects is of order $\mathcal{O}(\hbar^{0})$ and therefore cannot balance it out.  The GSL also seems to be violated semiclassically on apparent horizons \cite{FPSTap}.

\begin{figure}[hbt]
\centering
\includegraphics[width=.75\textwidth]{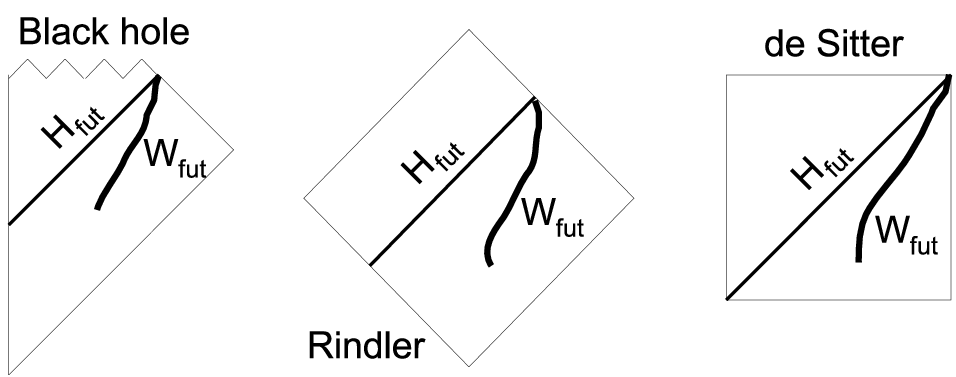}
\caption{\small{Black hole horizons, Rindler horizons, and de Sitter horizons are all special cases of ``causal horizons''.  The future causal horizon $H_\mathrm{fut}$ is defined as the part of the spacetime which is causally visible to some future-infinite timelike worldline $W_\mathrm{fut}$, shown as a thick line.  The GSL declares that the entropy is increasing with time on complete spatial slices outside of $H_\mathrm{fut}$ (shown as red lines).
}}\label{Wfut}
\end{figure}

Following Jacobson and Parentani \cite{JP}, I will assume that the GSL applies to the ``future causal horizon'' of any future-infinite timelike worldline $W_\mathrm{fut}$ (an ``observer'').  This causal horizon is defined as $H_\mathrm{fut} = \partial I^-(W_\mathrm{fut})$, the boundary of the past of the observer (Fig. \ref{Wfut}).

Given any two complete spatial slices $\Sigma$ and $\Sigma^\prime$ with the latter nowhere to the past of the former, the GSL then says that:
\begin{equation}
S_{\mathrm{gen}}(\Sigma^\prime\,\cap\,I^-(W_\mathrm{fut})) \ge S_{\mathrm{gen}}(\Sigma\,\cap\,I^-(W_\mathrm{fut})),
\end{equation}
where here the ``outside'' of the horizon used to define $S_{\mathrm{gen}}$ is taken to be the  side on which the observer is, even for cases like de Sitter space where the observer is enclosed by the horizon (see Fig. \ref{Wfut} for examples of pairs of  slices for which the generalized entropy increases.)

Note that by continuity $W_\mathrm{fut}$ may also be taken to be a lightlike ray whose affine parameter is infinite to the future, since there exist accelerating timelike worldlines which asymptotically approach any lightlike ray.  In this case $W_\mathrm{fut}$ may lie on its own horizon.

Just like the OSL, the GSL comes in two versions depending on whether we choose the fine-grained or coarse-grained definition of the state $\rho$ used to compute the entropy.  In the case of the fine-grained GSL, there can still be a nontrivial entropy increase due to the fact that information can fall across the horizon between $\Sigma$ and $\Sigma^\prime$.  (Another way of saying this is that even in the fine-grained picture we are still coarse-graining over all the information inside the horizon, a fully objective form of coarse-graining \cite{sorkin10theses}).  Since this is the only way entropy can change in the fine-grained picture, it follows that the only part of a spatial slice $\Sigma$ that matters is where it crosses the horizon.

By analogy to the fine-grained OSL, the fine-grained GSL ought to hold for every state of the universe, without needing to impose any initial condition.  (This can be explicitly checked for many of the existing proofs of the GSL in particular regimes \cite{10proofs, myproofs}.)  And if the GSL is true in all states, its time-reverse must also be true in all states \cite{anec}.\footnote{Technically the laws of physics are invariant under CPT, not T by itself.  This does not affect the argument because the generalized entropy is invariant under C and P.  But if the laws of quantum gravity were to violate CPT, the GSL and its time-reverse might be independent of each other.}
  
The $\overline{\mathrm{GSL}}$ states that for any past-infinite worldline $W_\mathrm{past}$, the past horizon $H_\mathrm{past} = \partial I^+(W_\mathrm{past})$ cannot \emph{increase} as time passes:
\begin{equation}
S_{\mathrm{gen}}(\Sigma^\prime\,\cap\,I^+(W_\mathrm{past})) 
\le S_{\mathrm{gen}}(\Sigma\,\cap\,I^+(W_\mathrm{past})).
\end{equation}
Of course, it does depend on the initial conditions whether there \emph{are} any past-infinite worldlines to which the $\overline{\mathrm{GSL}}$ might be applied.  If there are none then the $\overline{\mathrm{GSL}}$ is trivially true, although it may still be useful in astrophysical settings in which the spacetime may be treated as asymptotically flat and there are approximate $W_\mathrm{past}$'s.

On the other hand, the coarse-grained GSL would also take into account any entropy production of the matter outside the event horizon.  This has the advantage of treating ordinary thermodynamic processes on the same footing as the horizon thermodynamics, but has the disadvantage that the truth of the GSL must now depend on the existence of a low-entropy initial condition.  In particular the spacetime volume between $\Sigma$ and $\Sigma^\prime$ must have its thermodynamic arrow of time pointing to the future.  

Except for the discussion of the arrow of time in a past-infinite universe (section \ref{begin}), the results of this article will use only the fine-grained version of the GSL.

The GSL as I have defined it has been proven for semiclassical rapidly-chaging perturbations to stationary horizons, for free fields and/or Rindler horizons \cite{myproofs}.  Proofs are also available for classical spacetimes, and for semiclassical quasi-steady processes \cite{10proofs}.  Whether the GSL holds in a full theory of quantum gravity is obviously less certain (cf. section \ref{QG}).  And in the case of higher-curvature corrections to Einstein gravity, it is not yet known whether even a classical second law holds \cite{lovelock}, except in the special case of $f(R)$ gravity \cite{fR}.

\section{Generalized Thermodynamics Theorems}\label{theorems}

\subsection{Monotonicity Properties of the Generalized Entropy}

Suppose we have a region of spacetime $R$ which is well described by semiclassical gravity.  ``Semiclassical'' is a term with multiple meanings \cite{10proofs}, but I will take it to mean the following:

\textbf{Semiclassical Expansion:} A region R will be said to be semiclassical if its physics can be accurately described by a finite number of terms in an expansion controlled by $\hbar G / \lambda^2$, where $\lambda$ is the length scale of whatever quantum fields are relevant to the problem.  This is a bootstrapping procedure in which we start with (i) a fixed classical background metric, (ii) quantize matter fields and/or linearized gravitons on this background, (iii) allow these fields to infinitesimally perturb the background due to nonlinear gravitational effects, (iv) allow that perturbation to the background to affect the matter fields again, etc.  For simplicity we will hold $G$ and $\lambda$ fixed, and write the terms of the expansion with respect to $\hbar$.\footnote{This is distinct from the semiclassical approximation involving a large number $N$ of species, in which $N\hbar$ is held fixed as one takes the $\hbar \to 0$ limit, which will be discussed in section \ref{quantgeo}.}

For most purposes involving gravitational thermodynamics, it is sufficient to stop at step (iii), that step being needed only to calculate changes in the Bekenstein-Hawking entropy (which has an $\hbar^{-1}$ in the denominator).

We assume (without proof) that this procedure can be made well-defined using perturbative quantization of gravitons, which despite its nonrenormalizability should be valid as an effective field theory when treated using an ultraviolet cutoff much less than the Planck scale \cite{burgess}.  We assume that at finite orders in $\hbar$, this can be treated as if it were an ordinary quantum field theory with unitary evolution between Cauchy surfaces.  When $\hbar$ is infinitesimal the gauge symmetries of the graviton correspond to \emph{infinitesimal} diffeomorphisms.  This indicates that any observable $O$ of order $\hbar^n$ can be localized on the background spacetime up to terms which are higher order in $\hbar$.

In this semiclassical context, the generalized entropy will be assumed to take the following form:

\textbf{Generalized Entropy:} The generalized entropy of any codimension 2 surface will be assumed to take the form
\begin{equation}
S_\mathrm{gen} = \frac{A}{4 \hbar G} + Q + S_\mathrm{out},
\end{equation}
where $A$ is the expectation value of its area and $S_\mathrm{out}$ is the von Neumann entropy of the region spatially exterior to it, $G > 0$ is the value of Newton's constant at the renormalization scale, and the correction $Q$ to the Bekenstein-Hawking entropy is assumed to be subleading in $\hbar$ (or some other small parameter such as string length).  This is natural if $Q$ comes from radiative corrections, as described in section \ref{GSL}.

We assume that divergences in $S_\mathrm{out}$ can be regulated using some ultraviolet regulator such as the mutual information \cite{Imono}.  This regulator must cut off the entanglement entropy at distance scales less than some $\epsilon$, much smaller than the length scale $\lambda$ of the quantum fields (so as to capture all the convergent physics) yet larger than $L_\mathrm{Planck}$ so as to avoid the quantum gravity regime.

Because fine-grained entropy is conserved, all Cauchy surfaces of the exterior should have the same value of $S_\mathrm{out}$.

In the semiclassical regime one can show the following useful theorem about the increase of the fine-grained generalized entropy when comparing two null surfaces:

\textbf{Theorem 1:}  Let $N$ and $M$ each be future null surfaces of codimension 1, each of which divides spacetime into two regions, an ``interior'' $\mathrm{Int}$ and an ``exterior'' $\mathrm{Ext}$.  Let $M$ be either within or on $N$ everywhere (i.e. $M\,\cap\,\mathrm{Ext}(N)$ is empty).  (The location of the null surfaces might in general depend on the state of the fields.)

Let there be a null geodesic $g$ which lies on both $N$ and $M$, and a time slice $\Sigma$ which intersects $g$ at $g_0$.  Assume that in some neighborhood of $g_0$, the spacetime is semiclassical, and $N$ and $M$ are both smooth.\footnote{A typical null surface will develop cusps where its generators enter or leave the surface, and at these points the surface will not be smooth.  On a smooth spacetime these nonsmooth parts of the null surface are usually of lower dimension, so this assumption is reasonable if the point $g_0$ is generic.}  Very close to $g_0$, the null surfaces $N$ and $M$ will nearly coincide, but they may be separated by a small proper distance $f$.

For any neighborhood in the vicinity of $g_0$, there exists a way to evolve the time slice
$\Sigma$ forwards in time in that neighborhood to a new slice $\Sigma^\prime$, such that the generalized entropy increases faster on $M$ than on $N$:
\begin{equation}\label{ft}
\Delta S_\mathrm{gen}(\Sigma\,\cap\,\mathrm{Ext}(M)) - 
\Delta S_\mathrm{gen}(\Sigma\,\cap\,\mathrm{Ext}(N)) \ge 0.
\end{equation}
where $\Delta$ indicates the change in a quantity when evolving from $\Sigma$ by $\Sigma^\prime$.\footnote{Both $\Sigma$ and $\Sigma^\prime$ are assumed to be approximately constant over the length scale set by the proper distance $f$---otherwise one could satisfy Theorem 1 simply by evolving forwards in time on only one of the two null surfaces $N$ or $M$!}  Theorem 1 will be proven using a series of three Lemmas.

\begin{figure}[hbt]
\centering
\includegraphics[width=.6\textwidth]{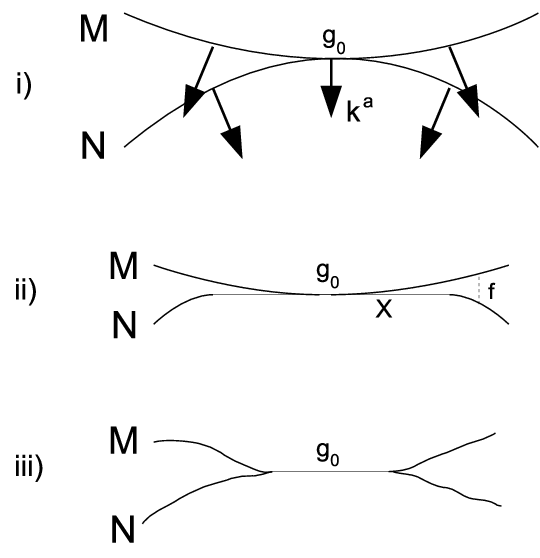}
\caption{\small{Two null surfaces $N$ and $M$ are pictured as they appear at one time, on the slice $\Sigma$.  $N$ is nowhere inside of $M$, and coincides with $M$ at $g_0$.  (i) the generating null vectors $k^a$, projected onto the surface $\Sigma$, must be normal to the null surfaces.  Because $M$ can only bend inwards relative to $N$ at $g_0$, it is expanding faster than $N$ there (Lemma A).  (ii) $f$ is the proper distance between the two null surfaces $N$ and $M$, viewed as a function of $M$.  Near the point $g_0$, $f$ is very gently sloped, and thus points on $N$ and $M$ may be identified.  Integration of $\nabla^2 f$ shows that it is always possible to find a point $X$ near $g_0$ at which $M$ is expanding faster than $N$, unless (iii) the surfaces coincide exactly in a neighborhood of $g_0$ (Lemma B).
}}\label{fronts}
\end{figure}

\textbf{Lemma A:} At the point $g_0$, the surface $M$ is expanding at least as fast as $N$ is.  See Fig. \ref{fronts}.

\textbf{Proof:} Since $N$ and $M$ coincide and are smooth at the point $g_0$, and $M$ cannot cross over from 
$\mathrm{Int}(N)$ to $\mathrm{Ext}(N)$, $N$ and $M$ must share the same tangent plane.  The null extrinsic curvature of one of the null surfaces, e.g. $N$ is defined as:
\begin{equation}\label{B}
B_{ab} = h^c_a h_{bd} \nabla_c k^d
\end{equation} 
where $h_{ab}$ is the pullback of the metric tensor onto the codimension 2 surface $\Sigma\,\cap\,N$, and $k^a$ is a (future-oriented) null vector pointing in the direction of the null generators on $N$.  

The null extrinsic curvature measures the change in the geometry of the null surface as it moves in the $k^a$ direction.  There are two contributions.  The first is a temporal component which arises when the slice $\Sigma$ itself has extrinsic curvature, but this may be disregarded because it is the same for both $N$ and $M$.  The second is a spatial component proportional to the extrinsic curvature $K_{ab}$  of $\Sigma\,\cap\,N$ in $\Sigma$ (with a normalization factor coming from the size of the projection of $k^a$ onto the slice $\Sigma$).  For any vector $v_a$ and point $x$, the extrinsic curvature component $K_{ab} v^a v^b (x)$ measures how much the surface $N$ curves away from its tangent plane, to second order, as one travels away from $x$ in the direction of $v^a$.  A positive value means that it curves away from the direction of motion of the null surface, and a negative value means that it curves towards the direction of motion.  

The fact that $M$ is inside of $N$ now places constraints on the extrinsic curvature of $N$ and $M$ at $g_0$.  Since $N$ is outside of $M$, $N$ must bend outwards by at least as much as $M$ does.  Hence:
\begin{equation}\label{bend}
B_{ab}^{(M)} v^a v^b \ge B_{ab}^{(N)} v^a v^b.
\end{equation}
The expansion of a null surface is related to the null extrinsic curvature as follows:
\begin{equation}\label{exp}
\theta \equiv \frac{1}{A}\frac{dA}{d\lambda} = B_{ab} h^{ab},
\end{equation}
where $A$ is the infinitesimal area near a generator, and $\lambda$ is an affine parameter satisfying $\lambda_{,a} k^a = 1$.  Eq. (\ref{bend}) then requires that in the neighborhood of $g_0$,
\begin{equation}\label{theta}
\theta^{(M)} \ge \theta^{(N)}.
\end{equation}
Q.E.D.

If the strict inequality $\theta^{(M)} > \theta^{(N)}$ holds, then by continuity inequality (\ref{theta}) also holds in a neighborhood of the point $g_0$.  In the classical limit $S_\mathrm{out}$ can be neglected, so $\theta$ gives the change of entropy.  In this special case, Theorem 1 follows.  In the saturated case where $\theta^{(M)} = \theta^{(N)}$, Lemma A is not enough.  In order to prove the classical version of Theorem 1, it is necessary to move a small distance away from the point $g_0$:

\textbf{Lemma B:} In any small neighborhood of $g_0$, either there is a point $X$ at which $\theta^{(M)} > \theta^{(N)}$, or else $M$ and $N$ coincide everywhere in that neighborhood.  In the former case, the area increases faster on $M$ than $N$ when $\Sigma$ is pushed forwards in time sufficiently close to the point $X$; in the latter case, the area increase is the same for $M$ and $N$ in the whole neighborhood.  Either way, Theorem 1 holds classically.

\textbf{Comment 1.1:} Lemma A is a special case of Lemma B, and was proven separately for pedagogical reasons.

\textbf{Proof:} On the spatial slice $\Sigma$, let the shortest proper distance between the surfaces $M$ and $N$ be given by a function $f(M)$.  Since the tangent planes of $N$ and $M$ coincide at $g_0$, $f$ vanishes to zeroth and first order as one moves away from $g_0$.  Since $f$ is only nonzero at second order and higher, in a neighborhood of lengthscale $\epsilon$, $f \apprle \epsilon^2 \ll \epsilon$.  Because the distance between $M$ and $N$ is in this sense small, it is possible to identify points on $N$ and $M$, permitting the function to be defined on either of the two null surfaces: $f(N) = f(M)$.

This identification of points on $N$ and $M$ also allows the null generating vectors $k^a$ to be compared on $N$ and $M$ (Fig. \ref{fronts}).  When the $k^a$ of $N$ or $M$ is projected onto $\Sigma$, it must be normal to that surface, because a lightfront always travels in the direction perpendicular the front itself.  So $k^a(\Sigma) = c n^a$, where $n^a$ is an outward pointing normal vector and $c > 0$ is an arbitrary constant depending on the normalization of the affine parameter $\lambda$ on $N$ and $M$.  In order to compare the $k^a$ vectors, $\lambda$ will be chosen so that $c = 1$ everywhere on $N$ and $M$.

For small $\nabla f$ this can be used to find the difference between $k^a$ on $N$ and $M$.
\begin{equation}
\Delta k^a = k^{a(M)} - k^{a(N)} = \nabla^a f + \mathcal{O}((\nabla f)^2),
\end{equation}
where up to the higher order terms, $\Delta k^a$ lies on the $D-2$ dimensional surface $M\,\cap\,\Sigma$ (or $N\,\cap\, \Sigma$).  The extrinsic curvature difference can now be calculated from Eq. (\ref{B}):
\begin{equation}
\Delta B_{ab} = B_{ab}^{(M)} - B_{ab}^{(N)} = \nabla_a \nabla_b f,
\end{equation}
where the covariant derivatives are intrinsic to the surface $\Sigma\,\cap\,N$.  Together with Eq. (\ref{exp}) this implies
\begin{equation}
\Delta \theta = \theta^{(M)} - \theta^{(N)} = \nabla^2 f,
\end{equation}
which is a total derivative.  Let $M$ (or $N$) be labelled by an $r$ coordinate representing the proper distance from $g_0$, and let $d\sigma$ be the volume element on the $(D - 3)$ dimensional space of constant $r$ on $M\,\cap\,\Sigma$.  

Let us define a Green's function $G(y)$ on the ball of points $y$ with $r < R$, to be the solution to these equations:
\begin{equation}
-\nabla^2 G(y) = \delta^{D-2}(y); \qquad G|_{r=R} = 0.
\end{equation}
For a sufficiently small $R$, the metric $h^{ab}$ is very close to being a flat Euclidean metric, so that $G \propto (r^{D-4} - R^{D-4})/(D-4)$ (or $\ln(R/r)$ in $D = 4$).  In any dimension, $G(y) > 0$ for $r < R$, and thus $\partial_r G|_{r=R} < 0$.  For sufficiently small $R$ these inequalities must continue to hold if the metric is slightly deformed by nonzero curvature.  One can now use $G$ to integrate $\Delta \theta$ on the codimension 2 ball $B$:
\begin{equation}\label{integral}
\int_B G\,\Delta \theta\,d^{D-2}y = \int_B G\,\nabla^2 f d^{D-2}y
= -\int_{\partial_B} f\,\partial_r G \,d\sigma\ge 0.
\end{equation}
where we have integrated by parts twice and used the fact that $f(0) = 0$.  

Now either (i) $f = 0$ in a neighborhood of $x$, or else (ii) one can find arbitrarily small values of $R$ such that the right hand side of Eq. (\ref{integral}) is strictly positive, in which case $\Delta \theta$ must also be positive for at least some points arbitrarily close to $x$.  Q.E.D.


The subject of the third lemma is the outside entropy term $S_\mathrm{out}$, about which nothing has yet been shown.

\textbf{Lemma C:} If the two surfaces $N$ and $M$ coincide in a neighborhood of $g_0$, and $\Sigma$ is evolved forwards in time to $\Sigma^\prime$ in this neighborhood, the entropy $S_\mathrm{out}$ is increasing faster on $M$ than on $N$.

\begin{figure}[phtb]
\centering
\includegraphics[width=.8\textwidth]{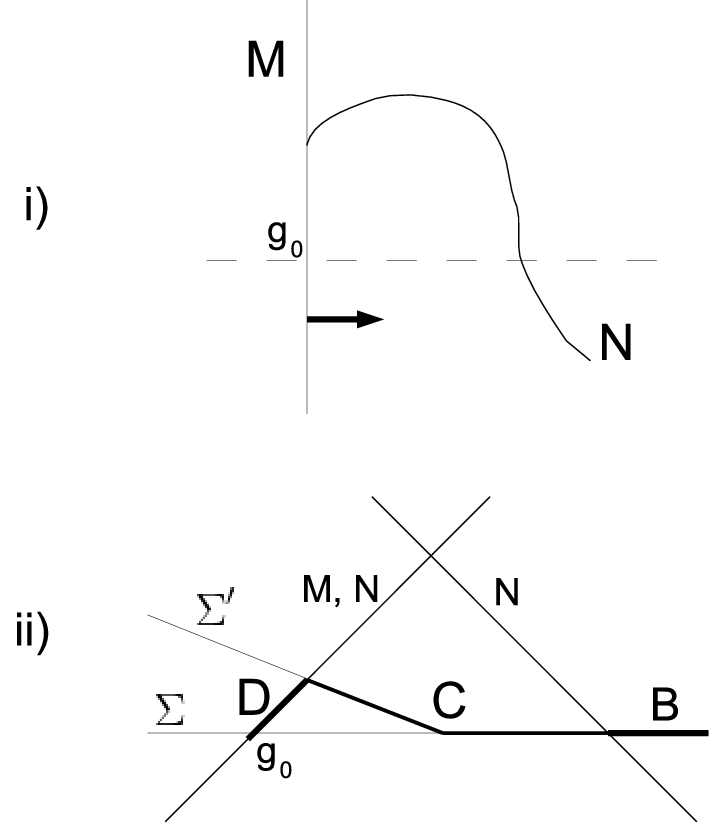}
\caption{\small{
(i) Two null surfaces $M$ and $N$ drawn on a time slice $\Sigma$, where $N$ is nowhere inside of $M$.  They coincide in a neighborhood of $g_0$.  This is the same situation as Fig. \ref{fronts}, illustrated with a different choice of $M$ and $N$.  The dotted line is the spatial cross-section used to make (ii) a spacetime diagram of the same situation.  $N$ and $M$ move outwards at the speed of light.  The time slice $\Sigma$ is evolved forwards in time to a new time slice $\Sigma^\prime$ near $g_0$.  All the information in $\mathrm{Ext}(M)\,\cap\,\Sigma$ is contained in three regions: $B$, $C$ and $D$.  Removal of the region $D$ can only decrease the amount of entanglement between $B$ and $C$, which can be used to show that the entropy outside of $M$ increases faster than the entropy outside of $N$.
}}\label{lemmaC}
\end{figure}

\textbf{Proof:} There is an information theoretical quantity called the mutual information $I(B,\,C)$, defined for any two disjoint systems $B$ and $C$, which measures the amount by which the entropy fails to be additive:
\begin{equation}\label{I}
I(B,\,C) = S(B) + S(C) - S(B\,\cup\,C).
\end{equation}
The mutual information measures the amount of entanglement between the systems $B$ and $C$.  For all quantum mechanical systems, this quantity is monotonically increasing as one increases the size of one of the systems by adding a third system $D$ \cite{Imono}:
\begin{equation}\label{mono}
I(B,\,C\,\cup\,D) \ge I(B,\,C).
\end{equation}
This makes sense intuitively, since one expects that the amount of entanglement between two systems can only be increased when one system is enlarged.  This property can be exploited by setting:
\begin{eqnarray}
B = \mathrm{Int}(N)\,\cap\,\mathrm{Ext}(M)\,\cap\,\Sigma, & \\
C = \mathrm{Ext}(N)\,\cap\,\Sigma^\prime, & \\
D = N\,\cap\,\Delta \Sigma, &
\end{eqnarray}
where $\Delta \Sigma$ is the spacetime volume between $\Sigma$ and $\Sigma^\prime$.  See Fig. \ref{lemmaC}.  

Now by Eq. (\ref{I}),
\begin{equation}\label{IBC}
I(B,\,C) = S(B) + S(\mathrm{Ext}(N)\,\cap\,\Sigma^\prime) - S(\mathrm{Ext}(M)\,\cap\,\Sigma^\prime).
\end{equation}
Similarly,
\begin{equation}\label{ID}
I(B,\,C\,\cup\,D) = S(B) + S(\mathrm{Ext}(N)\,\cap\,\Sigma) - S(\mathrm{Ext}(M)\,\cap\,\Sigma),
\end{equation}
where the slice $C\,\cup\,D$ has evolved backwards to the surface $\Sigma$, using the fact that unitary time evolution preserves the entropy.  Substituting Eq. (\ref{IBC}) and Eq. (\ref{ID}) into the monotonicity Eq. (\ref{mono}), one obtains
\begin{eqnarray}\label{Smono}
\Delta S_\mathrm{out}(M) \ge \Delta S_\mathrm{out}(N),
\end{eqnarray}
which shows that the outside entropy is increasing as fast for $M$ as for $N$.  Q.E.D.

\textbf{Proof of Theorem 1:} In the semiclassical limit, any effect which is higher order in $\hbar$ will be dominated by any nonzero effect which is lower order in $\hbar$.  Let the leading order contribution to $\theta^{(M)} - \theta^{(N)}$ be of order $\hbar^{p + 1}$, which 
corresponds to an order $\hbar^p$ contribution to $\Delta S_\mathrm{H}^{(M)} - \Delta S_\mathrm{H}^{(N)}$, since the Bekenstein-Hawking entropy (\ref{BH}) has an $\hbar$ in the denominator.  Lemma B says that at every order in $\hbar$, either $N$ and $M$ coincide or else $\Delta S_\mathrm{H}^{(M)} - \Delta S_\mathrm{H}^{(N)} > 0$ for an appropriate choice of $\Sigma$ evolution.  By applying Lemma B to order $\hbar^{p + 1}$, one obtains that the order $\hbar^p$ contribution to $\Delta S_\mathrm{H}^{(M)} - \Delta S_\mathrm{H}^{(N)}$ is positive.  By applying Lemma B at order $\hbar^p$, one obtains that $N$ and $M$ coincide at order $\hbar^p$.  Since $Q$ is subleading, there is no $\hbar^p$ order contribution coming from $\Delta Q^{(M)} - \Delta Q^{(N)}$.

Let the leading order contribution to $\Delta S_\mathrm{out}^{(M)} - \Delta S_\mathrm{out}^{(N)}$ be of order $\hbar^q$.  If $p \le q$, then the area term dominates over the entropy term.  If $p \ge q$, then since at this order the null surfaces coincide, Lemma C says that the $S_\mathrm{out}$ increases faster for $M$ than $N$.  Either way, Theorem 1 follows.

The only case not covered by the above argument is when both $p = q = +\infty$, i.e. when $N$ and $M$ coincide to all orders in $\hbar$.  But then their generalized entropy is identical to all orders, and thus Theorem 1 is true.  Q.E.D.

\textbf{Corollary 1.2:} At least semiclassically, one can extend the notion of a causal horizon to the boundary of the past of the \emph{union} of any number of future-infinite timelike or lightlike worldlines.  The reason is that any point on such a horizon must lie on the horizon of one of the worldlines, and then Theorem 1 shows that the GSL for the union is inherited from the GSL for that worldline.

\textbf{Corollary 1.3:} On the other hand, if one measures $S_\mathrm{out}$ in a region less than the whole exterior of a horizon, one does not always expect the entropy to increase.  In particular, in the Hartle-Hawking state, the existence of nonzero entanglement will make Eq. (\ref{mono}) positive, which implies that any region less than the whole exterior will have decreasing entropy, as in Ref. \cite{deriv}.

\textbf{Comment 1.4:} The semiclassical assumption is unnecessary so long as $M$ and $N$ coincide in a neighborhood of $g_0$.  That is because Lemma C depends only on purely information theoretical properties of $S_\mathrm{out}$, so it is only necessary to know that 
$\mathrm{Ext}(N)$ is a quantum subsystem of $\mathrm{Ext}(M)$.

\textbf{Comment 1.5:} With the possible exception of the no-warp-drive result in section \ref{warp}, the results in section \ref{app} only depend on the classical part of Theorem 1.  That is because in those cases, the null surface $M$ to which the theorem is applied already has a classical $\hbar^{-1}$ decrease in the generalized entropy, and the only thing which needs to be proven is that $N$ also has decreasing generalized entropy.

\begin{figure}[ht]
\centering
\includegraphics[width=.5\textwidth]{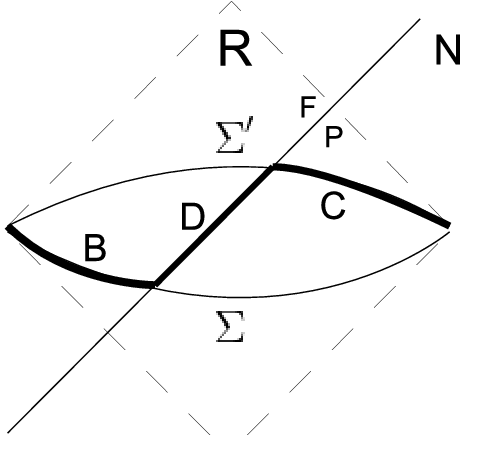}
\caption{\small{
The region $R$ is divided by a null surface $N$ into a past region $P$ and a future region $F$.  A time slice $\Sigma$ of $R$ is evolved forwards in time to $\Sigma^\prime$.  All information in $R$ is stored in the three regions $B$, $C$, and $D$.  Weak monotonicity implies that the generalized entropy of $F$ is increasing faster than the generalized entropy of $P$.
}}\label{thm2}
\end{figure}

\textbf{Theorem 2:} Let there be a globally hyperbolic region of spacetime $R$ cut across by a null surface $N$ into two regions $P$ and $F$, such that information can go from $P$ to $F$ by falling across $N$, but not vice versa.  Let $\Sigma$ and $\Sigma^\prime$ be two Cauchy surfaces of $R$, with the latter nowhere to the past of the former.  See Fig. \ref{thm2}.  Then the generalized entropy of $P$ minus the generalized entropy of $F$ cannot increase as time passes:
\begin{equation}\label{PminusF}
\Delta S_\mathrm{gen}(P) - \Delta S_\mathrm{gen}(F) \le 0
\end{equation}

\textbf{Proof:} The proof given in Ref. \cite{anec} is summarized here.  Since the regions $I^-(N)$ and $I^+(N)$ share the same boundary $N$ in the interior of $R$, the change in horizon entropy $\Delta S_\mathrm{H}$ is the same for both of them.  Furthermore any divergences in the entanglement entropy near the boundary must be the same on both sides \cite{anec}.  So the only quantity which may be different is the convergent part of the $S_\mathrm{out}$ terms.  In all quantum mechanical systems the entropy of three disjoint quantum systems $B$, $C$, $D$, obeys the weak monotonicity condition \cite{weakmono}:
\begin{equation}\label{weak}
S(B\,\cup\,D) + S(C\,\cup\,D) \ge S(B) + S(C).
\end{equation}
Intuitively, the more a system is entangled with one quantum system, the less it can be entangled with another.  Set $B = F\,\cap\,\Sigma$, $C = P\,\cap\,\Sigma^\prime$, and 
$D = N\,\cap\,\Delta \Sigma$ (where $\Delta \Sigma$ is the region between $\Sigma$ and $\Sigma^\prime$).  Unitary time evolution can be used to evolve the slice $B\,\cup\,D$ forwards in time onto $\Sigma^\prime$, and to evolve $C\,\cup\,D$ backwards in time onto $\Sigma$, without changing the entropy, so Eq. (\ref{weak}) evaluates to
\begin{equation}
S(F\,\cap\,\Sigma^\prime) + S(P\,\cap\,\Sigma) \ge S(F\,\cap\,\Sigma) + S(P\,\cap\,\Sigma^\prime),
\end{equation}
which then implies Eq. (\ref{PminusF}). Q.E.D.

\textbf{Corollary 2.1:} The (fine-grained) GSL and $\overline{\mathrm{GSL}}$ cannot hold on the same null surface $N$ unless they are both saturated, and weak monotonicity is also saturated.  In a suitably generic state, these inequalities will not be saturated, so $N$ cannot be both a past and a future horizon.

\textbf{Comment 2.2:} For Theorems 1 and 2, when applying the monotonicity properties (\ref{mono}) or (\ref{weak}), one may worry that the renormalization procedure needed to make $S_\mathrm{out}$ finite will interfere with the monotonicity property.  However, so long as the  infinite quantities subtracted off of the entropy of a region only depend on extensive, Lorentz invariant features of the region's boundary, the divergent terms combine in such a way as to cancel out of the final result.  Cf. Ref. \cite{anec} for a more detailed discussion of this point.

\textbf{Comment 2.3:} Because Theorem 2 follows from purely information theoretical properties of the entropy, the semiclassical approximation is not required.  The only requirements are that the regions $B$, $C$, $D$ be disjoint quantum systems with defined entropy, and that time evolution from $\Sigma$ to $\Sigma^\prime$ be unitary and causal.

\subsection{Quantum Trapped Surfaces}

The GSL, as formulated in section \ref{GSL}, applies only to causal horizons.  There always exist some null surfaces $N$ for which the generalized entropy is decreasing (e.g. for $N = \partial I^-(p)$, where $p$ is a point, the generalized entropy always decreases near $p$).  This does not contradict the GSL so long as $N$ is not a horizon.  The GSL is thus logically equivalent to the statement that any such null surface $N$ with decreasing entropy is not a causal horizon. 

That in turn means that there does not exist any worldline $W_\mathrm{fut}$ which is infinite to the future and for which $N$ is the boundary of the past of that observer.  This can be used to show that certain null surfaces must necessarily terminate:

\textbf{Theorem 3:}  Let $N$ be an achronal null surface, such that $g$ be a null generator of $N$, and let there be a point $g_0$ on $g$ at which the fine-grained generalized entropy is decreasing.  That is, there exists a spatial slice $\Sigma$, such that for any neighborhood around $g_0$, there is a way of pushing the slice forwards in time in that neighborhood to a new slice $\Sigma^\prime$, so that
\begin{equation}
\Delta S_\mathrm{gen}(\Sigma\,\cap\,\mathrm{Ext}(N)) \le 0.
\end{equation}
Thus if $N$ were a horizon, the fine-grained GSL would be violated for time evolution near $g_0$.

Suppose that the semiclassical approximation holds near $g_0$, while for the rest of the spacetime we assume only that it can be described by a Lorentzian manifold.  Then the GSL implies that the null generator $g$ cannot be extended infinitely on $N$ (either because it exits $N$, or because spacetime is null geodesically incomplete).

\textbf{Proof:} If $g$ stays on $N$ for an infinite affine distance, then it too must be achronal.  Furthermore it will have a horizon $H$ on which the GSL is satisfied, since the GSL must also apply to the horizons of infinite null rays as discussed in section \ref{GSL}.  $H$ must lie entirely on or to the past of $N$, because the past of $g$ must be a subset of the past of $N$.  Since $g$ is lightlike and achronal, $g$ must be a generator of $H$ as well as $N$.  Then Theorem 1 implies that the generalized entropy is also decreasing on $H$, which would violate the GSL.  Q.E.D.

If spacetime is globally hyperbolic, then a stronger result can be shown for certain surfaces.

\textbf{Global Hyperbolicity:} A spacetime is globally hyperbolic iff both a) there are no closed causal curves\footnote{Traditionally, global hyperbolicity requires also that the spacetime satisfy strong causality, but this apparently stronger form of global hyperbolicity was recently proven from the definition given here \cite{BernalSanchez}.} and b) for any two points $p$ and $q$, $J^+(p)\,\cap\,J^-(q)$ is compact.

Global hyperbolicity implies that there are Cauchy surfaces which intersect every timelike worldline exactly once.  Furthermore, one can find a smooth timelike vector field, whose integral curves must intersect any Cauchy surface exactly once \cite{HawkingEllis}.

\textbf{Quantum Trapped Surface:} Let there be a connected Cauchy slice $\Sigma$, containing a compact codimension 2 surface $T$ which divides it into two regions $\mathrm{Ext}(T)$ and $\mathrm{Int}(T)$, such that $\mathrm{Ext}(T)$ is noncompact.  Let a null surface $N$ be shot out from $T$ going outwards and to the future; $N$ may be defined more precisely as the future boundary of the domain of dependence of the exterior: $\partial^+ D[\mathrm{Ext}(T)]$.  Such a boundary is necessarily achronal.  If the fine-grained generalized entropy of $N$ is decreasing with time for each point $g_0$ on $T$ (in the sense described above in Theorem 3), then $T$ is a ``quantum trapped surface''.

\textbf{Comment 4.1:} In the classical limit, the generalized entropy becomes the area, and the definition reduces to the usual classical one: a surface $T$ is classically trapped if the area of the surface $N$ is decreasing everywhere at $T$ when moving outwards to the future.   By analogy to this, $T$ is quantum trapped if it is a compact surface for which the generalized entropy is decreasing everywhere on a compact, outward-moving $N$, near every point $g_0$ of $T$.

\textbf{Comment 4.2:} The existence of a quantum trapped surface does not necessarily violate the GSL, because the GSL only applies to future horizons, and $N$ is not necessarily a horizon.  However, if a quantum trapped surface \emph{were} a horizon, then it would violate the GSL everywhere on $T$.

\textbf{Theorem 4:}  Suppose there exists a globally hyperbolic spacetime with a quantum trapped surface $T$, as described above.  Let the semiclassical approximation be valid near $T$ (but not necessarily elsewhere).  Then the fine-grained GSL requires that the spacetime is not null geodesically complete, i.e. there is a singularity somewhere.

\textbf{Proof:} By Theorem 3, each of the null generator segments on $N$, i.e.~$\bar{g} \equiv g \cap N$, must terminate at some finite value of the affine parameter $\lambda$, because the generalized entropy is decreasing on it.  From this point on, the argument is the same as the classical Penrose singularity theorem \cite{HawkingEllis} which we summarize here:

Assume for contradiction that the manifold is null geodesically complete.  In that case, each segment $\bar{g}$ may be extended to the future beyond $N$, and therefore $\bar{g}$ includes its own endpoint, as part of $N$.  We can rescale the affine parameter so that $\lambda = 0$ at $T$ and $\lambda = 1$ at the endpoints.  This allows us to write $N$ as the topological product $T \times [0, 1]$, except that some of the endpoints at $\lambda = 1$ may be identified with each other.  Since $T$ is compact and so is the closed line segment $[0, 1]$, it follows that $N$ is compact.

However, global hyperbolicity prevents a noncompact spatial slice $\Sigma$ from evolving in time to a compact spatial slice $N$.  To see this, choose a smooth timelike vector field $t^a$ whose integral curves intersect $\Sigma$ once.  Since $N$ is achronal, the integral curves of $t^a$ intersect $N$ at most once.  $t^a$ can then be used to define a homeomorphism from $N$ to part of $\Sigma$.  Since $N$ is compact and without boundary, it must map to a subspace of $\Sigma$ which is itself compact and without boundary, but this contradicts the fact that $\Sigma$ is connected and noncompact.  Hence the spacetime must actually be null geodesically incomplete.  Q.E.D.

For more detailed descriptions of the Penrose proof, see Ref. \cite{HawkingEllis}.

\textbf{Comment 4.3:}  This shows that the Penrose singularity theorem can be generalized to quantum spacetimes so long as the fine-grained GSL holds.  This idea that the GSL gives rise to an analogue of trapped surfaces is implicit in the ``quantum Bousso bound'' proposed by Strominger and Thompson \cite{ST}.

\textbf{Comment 5:} Since by time-reversal symmetry the fine-grained $\overline{\mathrm{GSL}}$ must be just as true as the fine-grained GSL, the time-reversals of Theorems 1-4 also hold.

\section{Applications}\label{app}

\subsection{Black Holes and Babies}\label{black}

We will now apply the fine-grained GSL to the case of black hole collapse in order to show that there must be a black hole singularity somewhere (or else a Cauchy horizon due to failure of global hyperbolicity).  This requires a ``quantum trapped surface'' on which the GSL is being violated.  For a black hole with radius $r \gg L_\mathrm{Planck}$, the black hole should normally be described by an approximately classical metric.  So it is sufficient to find a surface $T$ which is classically trapped (i.e. its area is contracting before taking into account any quantum effects).  This decrease of area then implies an $\mathcal{O}(\hbar^{-1})$ decrease in the generalized entropy, which cannot be compensated for by the $\mathcal{O}(\hbar^{0})$ increase in $S_\mathrm{out}$.  Consequently the surface $T$ is also quantum trapped, which by Theorem 4 implies that it must be null geodesically incomplete, or else not globally hyperbolic---the exact same result obtained by the Penrose singularity theorem, but now applicable to certain quantum-mechanical situations.

However, just because there is a singularity somewhere does not necessarily mean that there must be a singularity everywhere.  Is it possible to avoid the singularities somehow and end up in a new universe?  Let us define more carefully what we mean by a baby universe: a baby universe is a spacetime region which is 1) inside the event horizon of a black hole, 2) contained in the future of the exterior of the black hole, and 3) can last for an indefinitely long proper time as measured by at least one observer.  That is, there must be able to exist a future-infinite worldline $W_\mathrm{fut}$ inside the event horizon.  (Note that if the baby universe ends up in a de Sitter type expanding phase, there may be multiple choices of $W_\mathrm{fut}$ which are separated by causal horizons.)

Proviso (1) ensures that the baby universe is distinct from the mother universe.  Proviso (2) rules out other universes which are not formed wholly from our own universe, but have pasts which are causally disconnected with our own universe.  In particular, the baby universe should not have come from an initial singularity of its own.  For example, Schwarzschild-de-Sitter would not be an example of a baby universe spacetime.  Proviso (3) is necessary to distinguish a baby universe from the usual picture of a black hole interior in which everything must end on a singularity in finite proper time.\footnote{Technically, this definition excludes baby universes which are eventually end in a Big Crunch without spawning any new universes themselves.  However, the no-go theorem might possibly be extended to such cases by arguing like this: As long as the baby universe expands for a long time without recollapse, there exist approximate $W_\mathrm{fut}$'s in the form of worldlines which exist for a very long time without collapse.  Since the GSL holds for infinite $W_\mathrm{fut}$, by continuity there ought to be some sense in which the GSL is very close to true for very long but finite worldlines.}

\begin{figure}[ht]
\centering
\includegraphics[width=.8\textwidth]{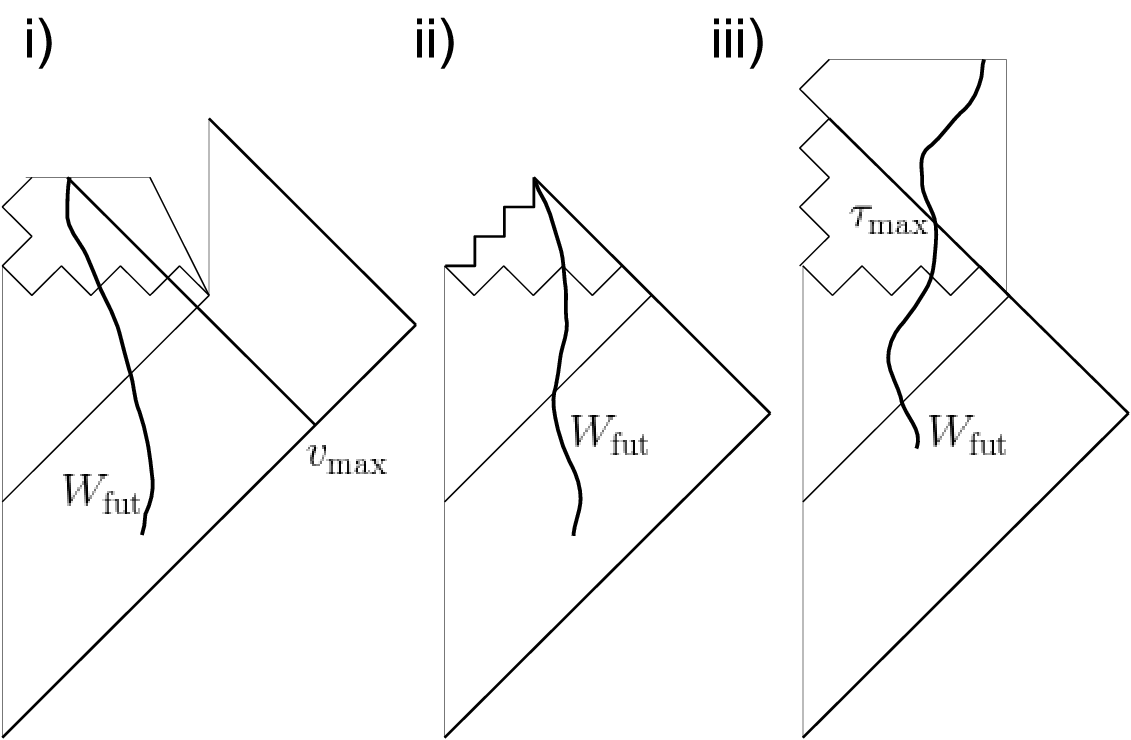}
\caption{\small{
Sample Penrose diagrams for baby universes forming from a collapsed black hole.  Each baby universe is shown to the future of a jagged line representing the classical singularity, and has a future infinite worldline $W_\mathrm{fut}$ falling into it.  This classical singularity may be resolved in some places by quantum effects, however the GSL requires that true singularities also appear somewhere in the spacetime (also shown by jagged lines).  It further places restrictions on the kinds of baby universes that are possible.  (i) A viable baby universe, ruled out by the GSL.  Since the baby universe ends in a de Sitter phase, several different future horizons can be selected depending on the choice of $W_\mathrm{fut}$.  In the example shown, the black hole evaporates completely, resulting in a disconnected space.  However, the GSL also excludes viable babies which remain connected to the mother universe.  (ii) A dependent baby universe, permitted by the GSL.  The black hole cannot evaporate completely.  (iii) A transcendental baby universe, ruled out by global hyperbolicity.  At $\tau_\mathrm{max}$ the worldline $W_\mathrm{fut}$ crosses a Cauchy horizon.  As it crosses it sees the entire history of the outside universe, infinitely blue shifted.
}}\label{babies}
\end{figure}

Assume that the black hole is in an asymptotically flat spacetime.  There are two different time coordinates which can be used to parametrize any given $W_\mathrm{fut}$.  Let there be a large stationary sphere around the black hole with a proper time coordinate $t$.  The advanced time coordinate $v(x)$ of any point $x$ is defined as the maximum value of $t$ on that part of the sphere which is to the past of $x$.  Another possible coordinate is the proper time $\tau$ of the timelike worldline $W_\mathrm{fut}$ inside of the baby universe.  Baby universes can be classified by means of the monotonic relationship between the two time coordinates $v(\tau)$ as follows (see Fig. \ref{babies}):

\begin{enumerate}
\item \textbf{Viable Babies:} Doctors call a fetus viable when it is capable of existing on its own without further life support from the mother's womb.  Adapting this definition for baby universes, let us define a ``viable baby universe'' as a baby which eventually becomes capable of causally existing on its own without support from the mother.  This requires that there exists \emph{at least} one $W_\mathrm{fut}$ such that as $\tau \to +\infty$, $v \to v_\mathrm{max}$ for some finite $v_\mathrm{max}$.  This is equivalent to saying that $W_\mathrm{fut}$ is in the future of a compact spacetime region.

\item \textbf{Dependent Babies:} A baby universe is dependent if it requires continued causal influences coming from the mother universe in order to remain in existence.  Apart from these causal influences, the baby can only last for a finite proper time.  This means that for all $W_\mathrm{fut}$, $\tau \to +\infty$, $v \to +\infty$.

\item \textbf{Transcendental Babies:} The remaining logical possibility is that for some $W_\mathrm{fut}$, $\tau \to \tau_\mathrm{max}$ for some finite $\tau_\mathrm{max}$, $v \to +\infty$.  This means that the baby universe requires an infinite period of gestation to reach a finite proper time, and then it goes on to become independent!  This very odd behavior violates global hyperbolicity, because if one takes a point $p$ outside the horizon, and a point $q \in W_\mathrm{fut}$ with $\tau > \tau_\mathrm{max})$, the region $I^+(p)\,\cap\,I^+(q)$ causally in between them is noncompact.  Thus $W_\mathrm{fut}$ crosses a Cauchy horizon at $\tau_\mathrm{max}$.  There would therefore be a failure of predictiveness across the Cauchy surface unless some new nonlocal physics principle were to come into play.  An example of such a spacetime is the analytically continued Reissner-Nordstr\"{o}m metric.
\end{enumerate}

\noindent A baby universe will start out being connected to the mother universe by an umbilical cord through the black hole event horizon.  However, a black hole left in a vacuum will eventually radiate away its mass until it reaches the Planck scale.  What happens after that depends on ones assumptions about quantum gravity.  If the black hole evaporates completely, then the connection must be broken at a finite value of $v$.  Call the resulting baby universe a \textit{disconnected} baby.  (We will assume that once the baby universe disconnects, it remains separate rather than reconnecting at a later time.)  Such topology changing spacetimes are non-globally-hyperbolic \cite{geroch}.  This kind of violation of global hyperbolicity might well be physically reasonable though, since the loss of predictivity only occurs at a single point in the manifold.  (Quantum gravity might restore predictivity, by specifying the dynamics of such topology-change points.)

The alternative is a \textit{connected} baby, which always remains connected to the mother universe, either because there is a remnant left over from black hole evaporation, or because the black hole is illuminated by sufficient infalling matter to prevent total evaporation.  It is also possible for a connected baby to be viable if e.g. it enters a de Sitter expansion phase, so that horizons form around individual future worldlines.

All viable babies---whether globally hyperbolic or not---are ruled out by the GSL, because the viability condition ensures that $W_\mathrm{fut}$ has a spatially-compact future horizon $\partial I^-(W_\mathrm{fut})$ existing outside the black hole, which separates the points which can causally influence the baby universe from the points which cannot.  However, when one traces the horizon $\partial I^-(W_\mathrm{fut})$ to the asymptotic region far from the black hole, then its generalized entropy must be decreasing with time.  To show this, we will first consider the spherically symmetric case, and then generalize to the non-spherically symmetric case.  We will assume that the semiclassical approximation is valid in the asymptotically flat region, far from the black hole.

In the spherically symmetric case, the future horizon is defined by $v = v_\mathrm{max}$ (the last moment of advanced time from which a signal can reach the baby universe).  Far from the black hole, the horizon is therefore a contracting sphere.  Since the area of such surfaces is decreasing on the classical black hole background metric, the Bekenstein-Hawking area entropy is shrinking by an $\mathcal{O}(\hbar^{-1})$ term.  Any quantum corrections coming from $S_\mathrm{out}$ are $\mathcal{O}(\hbar^{0})$, which is of lower order in the semiclassical expansion.  Consequently the generalized entropy of a future horizon is decreasing, contrary to the GSL.

If the spacetime is not spherically symmetric, then $\partial I^-(W_\mathrm{fut})$ may partly lie to the past of $v = v_\mathrm{max}$.  However, it is still true that $\partial I^-(W_\mathrm{fut})$ reaches to past null infinity, and that the asymptotic area of $\partial I^-(W_\mathrm{fut})\,\cap\,\mathcal{I}^-$ is infinite.  On the other hand, the area of a compact slice of $\partial I^-(W_\mathrm{fut})$ is finite when the slice is taken at large (but finite) distance from the black hole.  In order to go from infinite to finite area, there must exist some point $g_0$ far from the black hole where $\partial I^-(W_\mathrm{fut})$ is contracting classically.  But then, by the same argument as in the preceding paragraph, the generalized entropy decreases at $g_0$.

Therefore if the GSL is true, no such $W_\mathrm{fut}$ can exist; in other words there is no viable baby universe.  This is a generalization of a theorem in classical general relativity using the null energy condition \cite{obstacle}.


So far we have not assumed that the baby spacetime is globally hyperbolic.  Global hyperbolicity would rule out the transcendental babies, as well as disconnected babies.  But of the two, transcendental babies seem much more pathological because of the ``infinite blueshift'' of the field modes falling across the horizon at late times.  So even if topology changing events are allowed, it still seems reasonable to disallow transcendental babies.  That would exclude the proposed quantum tunneling process, described in Ref. \cite{monopole}, as well as the baby universe spacetime of Ref. \cite{FMM}.  (Even if transcendental babies were allowed, they would probably be very sensitive to the long term fate of the universe, since they can only arise inside of black holes that have a finite probability of never evaporating completely.)

Thus, assuming both the GSL and global hyperbolicity, only dependent babies are permitted as a possibility.  The reason why the GSL does not forbid dependent babies is that $I^-(W_\mathrm{fut})$ includes the entire exterior of the black hole.  Thus there is no horizon separating the points which can influence the baby from the points that cannot.  However, because any babies must be dependent, the opportunities for universe creation are limited.

First of all, a dependent baby universe must always remain connected by its umbilical cord to the mother universe lest it die.  But any black hole that is left to itself will evaporate due to Hawking radiation.  So unless remnants are allowed, everything inside of the black hole must be destroyed if the black hole ever stops being fed.

Secondly, a dependent baby universe cannot have a cosmology similar to our own universe, which appears to be heading into a de Sitter expanding phase.  De Sitter space has a compact future horizon around any future-infinite worldline $W_\mathrm{fut}$.  Let there be a spatial slice $\Sigma$ on which $\Sigma\,\cap J^-(W_\mathrm{fut})$ is a compact region.\footnote{$J^-$ has been used instead of $I^-$ in order to make this region closed.}  For any non-transcendental baby, $v$ must be finite at every point in $\Sigma\,\cap\,J^-(W_\mathrm{fut})$.  Since $v$ is a continuous function, by compactness, this means that $v$ has an upper bound in the region $\Sigma\,\cap\,J^-(W_\mathrm{fut})$.  This means that no information can reach $W_\mathrm{fut}$ after a certain advanced time $v$; hence the baby universe is viable.  Consequently, no dependent baby universe can end up in a de Sitter expanding phase.

So the GSL requires that any baby universes (and their progeny forever) must remain dependent on this one, and have a different cosmology than our own universe appears to.  (Additional constraints on universe formation will be given by the $\overline{\mathrm{GSL}}$ in the next section.)

\paragraph{Restarting Inflation.} Assuming the GSL and global hyperbolicity, the same argument that rules out baby universes also implies that one cannot restart inflation in an asymptotically Minkowski space, since the compact future de Sitter horizons would violate the GSL in the Minkowski region of spacetime.  In other words, any inflationary region would become a viable baby universe, and would therefore be ruled out.  This corresponds to a classical result using the null energy condition \cite{obstacle}.  It is also in agreement with the AdS/CFT argument of Ref. \cite{myers}.  On the other hand, the prohibition on baby universes appears to conflict with semiclassical instanton calculations \cite{bubble} of quantum tunnelling probabilities to restart inflation.  However, these instanton calculations correspond to Euclidean manifolds over degenerate metrics.  Their validity is controversial \cite{banks}.

\paragraph{Traversable Wormholes.} Similarly, there can be no traversable wormholes between two distant regions of spacetime, because any worldline which crossed from past null infinity of one region, to future null infinity of the other, would have to have a classically contracting future horizon in the first region.  Again, this result is analogous to a classical result using the null energy condition \cite{worm}.

\paragraph{De Sitter and Anti-de Sitter} The arguments in this section can also be applied to asymptotically anti-de Sitter spacetime, and more generally to any spacetime in which large ingoing null surfaces are contracting.  (Thus it does not matter that Anti-de Sitter space violates global hyperbolicity due to its boundary at spatial infinity).  In the case of wormholes between two asymptotically AdS spacetimes, the prohibition of wormhole traversal is in accordance with AdS/CFT \cite{GSWW}.  Since such spacetimes would have two disjoint conformal boundaries, there is no way that information could be causally transmitted from one CFT to the other.

In the case of de Sitter space, ingoing null surfaces are contracting only if they are sufficiently small.  Therefore, the GSL only restricts baby universes and inflationary regions in de Sitter space if their horizons are at a distance scale shorter than the de Sitter radius.  It may be that this places constraints on eternal inflation scenarios.  However, in order to address eternal inflation it is necessary to carefully consider the role of entropy fluctuations (cf. section \ref{entfluc}, footnote \ref{inflation}).

\subsection{Big Bangs and Beginnings}\label{begin}

Did the universe have a beginning in time?  We have already discussed two incomplete arguments that it did:  a) In section \ref{OSL}, the coarse-grained OSL was used to argue that if the thermodynamic arrow of time always points forwards, there can only be a finite amount of entropy production in our past.  However, the argument failed for gravitational systems such as general relativity because of the possibility that the entropy might be able to increase without bound, permitting systems with infinite yet increasing entropy.  b) In classical general relativity, one can instead use the Penrose singularity theorem to argue that if the universe is spatially infinite, there must have been an initial singularity.  But this theorem uses the null energy condition, which fails for quantum fields.\footnote{There are also singularity theorems which apply to spatially finite universes \cite{HawkingEllis}, but these theorems use the strong energy condition.  This condition can be violated even by classical, minimally-coupled scalar fields (and was violated in the early universe, if inflationary cosmology is true).  There is little reason to believe that either these theorems or some quantum analogue apply to the early universe.}

In this section we will use generalized thermodynamics to prove quantum analogues of both (a) and (b).  Let us assume that although the early universe may have been quantum and inhomogeneous, at late times and at large-distance scales, the universe is described by some expanding classical Friedmann-Robertson-Walker (FRW) cosmology.  The application of the classical Penrose theorem to the Big Bang cosmology uses the fact that a sufficiently large sphere $T$ in an expanding FRW cosmology is an anti-trapped surface, i.e. even the inward moving null surface generated from $T$ is expanding.  When these rays are instead traced backwards in time, they are contracting and outwards moving.  Assuming global hyperbolicity, at least one ray must be null geodesically incomplete, which implies a singularity.

To generalize this result to the quantum case, note that if the anti-trapped surfaces are in a semiclassical region of spacetime, the fact that $T$ is classically anti-trapped means that it is also quantum anti-trapped, meaning that the generalized entropy of this past horizon decreases when one goes to the past.  Using the fine-grained $\overline{\mathrm{GSL}}$, it follows from Theorem 4 that the spacetime has an initial singularity, if spacetime is globally hyperbolic and space is noncompact.  It does not matter if the FRW cosmology has small inhomogeneous perturbations because small perturbations cannot eliminate the anti-trapped surfaces.

Just as in the black hole case, the mere fact that there is an initial singularity, does not necessarily tell us that there are no past-infinite worldlines $W_\mathrm{past}$ which avoid the singularity.  But some constraints can be placed on this possibility by assuming that there is such a $W_\mathrm{past}$ and then applying the fine-grained $\overline{\mathrm{GSL}}$ to it.  Suppose that a past horizon 
$\partial I^+(W_\mathrm{past})$ exists in the present day universe.  Because of the expansion of the universe, such a horizon should now be a large, nearly classical object.  And by the $\overline{\mathrm{GSL}}$, it must be nonexpanding everywhere.  In an expanding FRW cosmology, this is only possible if each connected component of the past horizon is compact and sufficiently small.  Therefore the entirety of the infinite universe would share a common past history, except for possibly a set of bubbles each with finite spatial volume.\footnote{By the generic condition used in the Corollary 2.1, these bubbles are bounded by future-trapped surfaces.  By Theorem 3, they must eventually contract to nothing.  If the FRW cosmology has no final singularity, this must happen simply by the lightrays crossing each other.  Therefore, there exists a complete spatial slice $\Sigma$ in the FRW cosmology with the property that the entire $\Sigma$ lies to the future of any past infinite worldline $W_\mathrm{past}$.}

This is in stark contrast with the standard hot Big Bang FRW cosmology, in which sufficiently distant spatial regions have never been in casual contact with one another.  But it is not too different from inflationary cosmology, in which the exponential expansion of the universe causes distant regions to share a causal past.  One way to provide spacetime with the requisite property would be if there were an infinite period of inflation to the past of the infinite universe.  Such a spacetime would have past-infinite worldlines, but would also be null geodesically incomplete due to the fact that an infinite inflating universe occupies only a piece of de Sitter space.

Similarly a spatially finite universe can easily have all points in its FRW phase eventually be in causal contact, and it does not even need any initial singularities to do so.  An example would be a $\Lambda$-FRW cosmology with a bounce.

In order to analyze the thermodynamic properties of such a past-infinite model, we now invoke the coarse-grained GSL.  Assuming in accordance with current observations that the universe will end up in a de Sitter-like expanding phase, there will be future-infinite worldlines $W_\mathrm{fut}$ beginning on Earth which will end up being surrounded by a compact future horizon, containing a finite amount of generalized entropy (approximately $A/4\hbar G$ of the horizon).  Assuming global hyperbolicity, this future horizon must remain compact as it is taken to the past, and includes in particular everything in the past of Earth.  This means that either 1) the generalized entropy inside the horizon has increased from arbitrarily negative values, or 2) there are only a finite amount of entropy producing processes in our past lightcone, or 3) the thermodynamic arrow of time is reversed somewhere in our past, so that the coarse-grained GSL does not hold.

Option (1) is not possible if the generalized entropy has a direct state counting interpretation in terms of discrete 
Planck-scale degrees of freedom, and in any case seems somewhat bizarre.  Option (2) would involve the universe being in a near-equilibrium state for the first ``half'' of eternity and then for some inexplicable reason exiting this equilibrium.  Option (3), although strange seeming, arises naturally if the low entropy ``initial conditions'' of the universe are actually imposed on some finite time slice (cf. section \ref{OSL}).  In this scenario, the universe can be said to have a beginning in a thermodynamic sense even if it does not have a beginning in a geometrical sense.

Option (3) works best if the universe is spatially finite.  In a spatially finite universe, the horizon of $W_\mathrm{fut}$ can, when traced backwards in time, intersect itself and disappear entirely at a time $t_*$.  (In the case of a spatially-infinite, globally hyperbolic universe the horizon can only disappear entirely by hitting an initial singularity.)  Since before $t_*$ there is no horizon, the fine-grained generalized entropy is simply a constant, equal to the fine-grained entropy of the total universe.  This explains why the fine-grained entropy does not decrease indefinitely when one goes backwards in time.  But if the horizon goes back in time forever, one would run into problems with the fine-grained GSL.  Since the fine-grained GSL does not depend on an arrow of time (cf. section \ref{GSL}), one would have to endorse options (1) or (2) with respect to the fine-grained generalized entropy, eliminating the benefit obtained from reversing the arrow of time.

Putting all these considerations together, if the GSL is a valid law of nature, it strongly suggests that either the universe had a finite beginning in time, or else it is spatially finite and the arrow of time was reversed previous to the Big Bang.\footnote{In a bouncing scenario, it is not logically necessary that the moment of lowest entropy was the same as the moment of smallest size, but it seems natural to make this identification.}  In the latter case, it could still be said that the universe had a beginning in a thermodynamic sense, because both branches of the cosmology would be to the thermodynamic future of the Big Bang.

\subsection{Warp Drives and Negative Mass Objects}\label{warp}

The fine-grained GSL can also be used to rule out certain kinds of warp drive spacetimes.  We will consider spacetimes $M$ which are asymptotically flat and globally hyperbolic, but which have in their interior some gravitational fields which are capable of bending the lightcones so as to enable superluminal travel.  See Ref. \cite{warps} for discussions of such spacetimes.

Since the positions of the lightcones depends on the choice of coordinates, it is necessary to provide a diffeomorphism-invariant definition of superluminal travel.  Coming up with a sensible definition is tricky, since diffeomorphisms can move around the start and finish points.  In order to deal with this, this section will only consider the case of \textit{asymptotic} warp drives, which speed up the propagation of light rays relative to the asymptotic structure of the spacetime.  

The approach here is inspired by Ref. \cite{olum}, which defined a lightray as being superluminal if it travels between a certain pair of 2-surfaces faster than any nearby lightray.  However, instead of looking at travel between two 2-surfaces separated by a finite distance, I will consider a lightray $g$ travelling from the past conformal boundary $\mathcal{I}^-$ to the future conformal boundary $\mathcal{I}^+$.

\begin{figure}[ht]
\centering
\includegraphics[width=.7\textwidth]{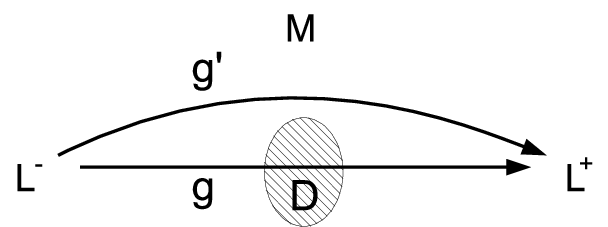}
\caption{\small{
This diagram shows light signals travelling through space from an asymptotic past origin $L^-$ to an asymptotic future destination $L^+$.  A warp drive is defined as a compact region $D$ of spacetime, such that some null curve $g$ passing through $D$ is advanced by a finite time relative to any curve $g^\prime$ which does not pass through $g$.
}}\label{warp1}
\end{figure}

Intuitively, a spacetime is a warp drive if there exists a compact region $D \in M$ such that light-speed signals which pass through $D$ are advanced by a finite time relative to signals which ''go around'' in $M - D$.  (See Fig. \ref{warp1}.)

\begin{figure}[ht]
\centering
\includegraphics[width=.9\textwidth]{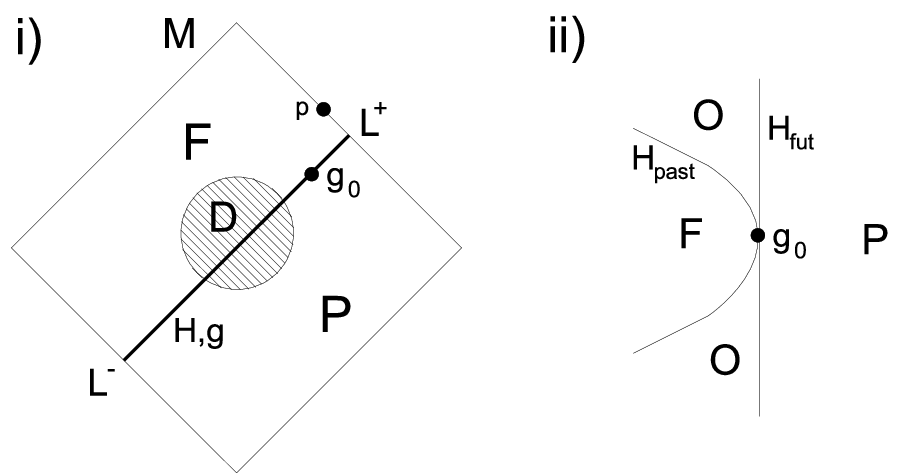}
\caption{\small{
i) A spacetime diagram of the warp drive.  The null curve $g$ passes through the region $D$ in order to connect points $L^-$ and $L^+$ on the conformal boundary.  There exist points $p$ to the null future of $L^+$ which cannot be accessed from $L^-$ without passing through the region $D$.  The null curve $g$ lies on both the future horizon $H_\mathrm{fut} = \partial I^-(L^+)$ and the  past horizon $H_\mathrm{past} = \partial I^+(L^-)$.  However, these horizons need not coincide except at $g$.  ii) A picture of a spatial slice that cuts through the point $g_0$.  The two horizons define three regions, F to the future of $g$, $P$ to the past of $g$, and the spacelike separated other region $O$.  Assuming the semiclassical approximation is valid at $g_0$, Theorems 1 and 2 can then be applied to show a violation of either the GSL or $\overline{\mathrm{GSL}}$.
}}\label{warp2}
\end{figure}

More precisely, let us define a ``warp drive'' as a compact spacetime region $D \in M$, with the property that there exist points $L^- \in \mathcal{I}^-$ and $L^+ \in \mathcal{I}^+$ such that (see Fig. \ref{warp2}):
\begin{enumerate}
\item $L^-$ and $L^+$ are achronal (i.e. they are not connected by any timelike curve), 
\item There exists a null curve $g$ travelling from $L^-$ to $L^+$ passing through the region $D$, but 
\item In the partial spacetime $M - D$, any null curves travelling from $L^-$ to $\mathcal{I}^+$ are delayed by a finite time.  In other words, for any other point $p \in \mathcal{I}^+$, if $p$ is null separated from $L^+$, and if $p$ is sufficiently close to $L^+$, then it is not possible to send a signal from $L^+$ to $p$. 
\end{enumerate}
To summarize this definition, $L^+$ and $L^-$ are lightlike separated in $M$, but spacelike separated on $M - D$.  This indicates that null curves which travel through $D$ are advanced by a finite time compared to those which go around $D$.

A few comments on this definition: First, conditions 1 and 2 implicitly require $g$ to be an achronal null curve, which in turn implies that it is a ``fastest possible'' geodesic connecting $\mathcal{I}^-$ and $\mathcal{I}^+$.  The existence of a curve with maximum possible speed follows from global hyperbolicity for any warp drive spacetime, since the space of causal curves between any two compact subsets of a globally hyperbolic spacetime is itself compact \cite{SW96}.  

The points $L^-$ and $L^+$ must be in diametrically opposite spatial directions, since all other points in $\mathcal{I}^+$ are chronal to any $L^-$.  This is why condition 3 above restricts attention to points in $\mathcal{I}^+$ which are null separated to $L^+$.

One might na\"{i}vely that even in flat spacetime, a region $D$ could satisfy this definition simply by blocking the shortest path between $L^-$ and $L^+$, and thus forcing any lightrays connecting $L^-$ and $L^+$ in $M - D$ to go around a longer way.  However, because $L^-$ and $L^+$ are infinitely far away, lightrays only need to bend by a small amount to get around $D$, leading to a delay that can be made arbitrarily small, meaning that condition 3 is not satisfied.

Finally, this definition applies only to warp drives that lead to a finite advance for lightrays travelling over an infinite distance.  The result does not apply to cases where there is a speed up only over a finite distance.

We will now show a contradiction between generalized thermodynamics and the existence of warp drives as defined above.  Since there exist infinite worldlines beginning on $L^-$ or ending on $L^+$, these points define Rindler-like horizons cutting through the spacetime $M$.  These two horizons cut $M$ into three regions, $P = I^-(L^+)$, $F = I^-(L^+)$, and the remainder $O = M - F - P$ ($P$ and $O$ cannot overlap or $L^-$ and $L^+$ would be timelike rather than lightlike).  $g$ is required to lie in both the causal past region $J^-(L^+)$ and the causal future region $J^+(L^-)$.  However, it cannot lie in the interior of either region, or there would be a timelike curve going from $L^-$ to $L^+$.  Consequently it must be a null geodesic lying on the boundary of these regions, on both the future horizon $H_\mathrm{fut} = \partial I^-(L^+)$ and the past horizon $H_\mathrm{past} = \partial I^+(L^-)$.  Assuming that $F$ obeys the GSL and $P$ obeys the $\overline{\mathrm{GSL}}$, it is now possible to derive a contradiction.

Although the fields may not be semiclassical in the region $D$, there ought to exist at least one point $g_0 \in g$ on which the semiclassical limit does hold in the asymptotically flat region away from any significant matter excitations.  At this point $g_0$, Theorem 1 may be invoked to show that if the GSL holds on $F$, it also holds on $F + O$.  Furthermore, Corollary 2.1 (choosing $R = M$) implies that if the GSL holds on $F + O$ and the $\overline{\mathrm{GSL}}$ holds on $P$, then the state is not generic.  But if there are any warp drive spacetimes, there are also generic ones, since if a lightray is advanced by a finite time by travelling through $D$ (condition 3), an infinitesimal generic perturbation of the spacetime must preserve this property.  It follows that either there are no (asymptotically flat, globally hyperbolic) warp drive spacetimes, or else the GSL or $\overline{\mathrm{GSL}}$ are violated.

This result is a generalization of the classical theorems of Refs. \cite{olum, supercensor, GW00}.

\paragraph{Positive Energy Theorem.}  An immediate corollary is that the GSL and $\overline{\mathrm{GSL}}$ together imply a positive energy theorem, by the principles outlined in Ref. \cite{SPW}.  Suppose we have an asymptotically flat spacetime containing an isolated compact object with some ADM mass.  Now an object with a positive ADM mass causes lightrays passing through its asymptotic gravitational field to be delayed.  For a negative mass object this Shapiro ``delay'' is actually an advance.  Thus an object with negative mass could be used for superluminal communication, as a type of warp drive.  Hence it is forbidden by the previous result.

The classical theorem \cite{SPW} has an important limitation which should be noticed: it requires the spacetime to be asymptotically flat at \emph{null} infinity, not just spacelike infinity.  In other words, the spacetime must remain sufficiently stable that its mass can be probed by a lightray going from past null infinity to future null infinity.  For example, there exist Kaluza-Klein spacetimes with negative ADM mass in which a ``bubble of nothing'' contracts and then expands, asymptotically approaching the speed of light \cite{witten82}.  This is not a contradiction because the bubble of nothing hits null infinity and prevents it from being asymptotically flat.  Similar problems arise for bubbles of AdS space inside of a ``false vacuum'' with zero cosmological constant.

For asymptotically Schwarzschild solutions in 4 dimensions, one can also derive a partial converse result: any solution with positive mass is \emph{not} a warp drive spacetime.  Because the gravitational potential falls off with distance like $1/r$, the integrated time delay is logarithmically divergent, and therefore the Shapiro delay from the asymptotic gravitational field is $+\infty$.  This overcomes any finite Shapiro advance coming from the interior of the spacetime.   Unfortunately, this makes the no-warp-drive result somewhat trivial in this case, since it does not rule out any asymptotically Schwarzschild solutions with positive mass.

However the result is not as trivial for $D \ge 5$ spacetimes, where the Shapiro delay for Schwarzschild is finite.  Nor is it trivial for asymptotically Anti-de Sitter spacetimes, which we consider next.

\paragraph{Causality in Anti-de Sitter.}  The results above can be directly generalized to the case in which the spacetime is asymptotically AdS, by choosing $L^+$ and $L^-$ to lie on the AdS boundary.  The proof of the result is analogous, notwithstanding the fact that anti-de Sitter space technically violates global hyperbolicity.  (The only reason why Theorem 2 needed global hyperbolicity was to ensure that the spacetime region had good causality properties.  But AdS space has equally good causality properties after one imposes boundary conditions at spatial infinity.)  This generalizes the classical theorems of Ref. \cite{woolgar94, GW00}.

This also fits in nicely with what is known about the AdS/CFT conjecture.  In order for a theory of gravity to have a field theory dual living on its boundary, it is essential that between two spatial locations $A$ and $B$ on the boundary, it is impossible to get from $A$ to $B$ any faster when travelling through the bulk, than when going around on the boundary.  Otherwise, it would be possible to send signals faster than light in the CFT \cite{PSW02}.  The GSL is a plausible physical principle enforcing this requirement.

\subsection{Time Machines}\label{time}

The final application of the fine-grained GSL will be to rule out time machines, which is again a generalization of classical results \cite{noctc}.  Up until now, we have assumed that spacetime is globally hyperbolic.  But global hyperbolicity rules out closed timelike curves (CTC's) by definition, making any proof trivial.  So in this section, we will assume instead that a) all CTC's in the spacetime are to the future of some point $p$, which is in turn to the future of $\mathcal{I}^-$, b) for any two points $p$ and $q$, $J^+(p)\,\cap\,J^+(q)$ is compact (i.e. the other component of global hyperbolicity \cite{BernalSanchez}), and c) spacetime is asymptotically flat.
  
The goal will be to show that a CTC can never form.  Since a CTC is an infinite worldline wrapped around the same points periodically, $\partial I^-(\mathrm{CTC})$ is a future horizon.\footnote{If you think that a circular worldline should not count as ``infinite'' in the relevant sense, simply consider a slightly wiggly line near the CTC which never exactly intersects itself.}  Excluding the future of $p$, spacetime is globally hyperbolic, so there is no problem defining complete slices for the generalized entropy.  By applying condition (b) to the point $p$ and any point on the CTC, $\partial I^-(\mathrm{CTC})$ must have compact slices on the globally hyperbolic part of the manifold.  Thus the generalized entropy should increase towards the future on $\partial I^-(\mathrm{CTC})$ on a compact slice.

But by asymptotic flatness, this compact horizon must be contracting at early times, violating the GSL for the same reason that baby universes did in section \ref{black}.  Consequently, assuming the GSL and condition b (which is the other half of global hyperbolicity), it is impossible for any experimenter sitting at a point $p$ in an asymptotically flat universe to arrange for a CTC to form.  Similarly, the time-reverse of the above argument using  $\overline{\mathrm{GSL}}$ shows that if there are currently existing time machines it is impossible for them to be destroyed.

\section{Does it still work for Quantum Gravity?}\label{QG}

The above results have been proven on the assumption that spacetime can be approximated by a smooth, globally-hyperbolic, Lorentzian manifold subject to a small quantum perturbation, such that the resulting spacetime satisfies the GSL exactly.  The question is whether we expect the result to hold even if some of these assumptions are relaxed.  In particular, the following questions arise:

What about thermodynamic fluctuations which cause the entropy to temporarily decrease, thus making the GSL not exact?  Do quantum fluctuations in the metric make the GSL ill-defined in the quantum gravity regime?  And what about global hyperbolicity and the other assumptions involved in the no-go results of section \ref{app}?

\subsection{Entropy Fluctuations}\label{entfluc}

All thermodynamic systems have fluctuations, since it is always possible that the degrees of freedom in a complex system will, by chance, temporarily enter an unlikely configuration.  When the entropy is defined as the Boltzmann entropy $S = \ln\,N$, where $N$ is the number of microstates in a macrostate, this can lead to a temporary decrease in the entropy.  However, in the approach to the second law which is used here, the entropy of quantum fields is defined in Eq. (\ref{von}) using the Gibbs entropy $S = -\mathrm{tr}(\rho\,\ln\,\rho)$.  As shown in section \ref{OSL}, this entropy can be proven to be exactly nondecreasing.  In the case of the GSL one also has the area term.  I have argued elsewhere \cite{10proofs} that one can take a similar interpretation of the generalized entropy if one defines the GSL using the expectation value $\langle A \rangle$, as suggested in Ref. \cite{fluct}.  The GSL as defined in this way need not have any downward fluctuations, which is convenient for proofs of the GSL.

Nevertheless, changing the definition of the entropy cannot change the underlying physics, and the entropy fluctuations are still present and physically important.  In the Gibbs interpretation, these entropy fluctuations appear when one invokes the probability interpretation of the density matrix $\rho$.  For example, consider the spin of an electron which has two states,up and down.  If the system is in uniformly mixed density matrix with diagonal $(1/2,\,1/2)$, the entropy is $\ln\,2$.  But this density matrix only represents our ignorance; the electron may well actually be in the up state.  And it is easy to show that the maximum entropy associated with any pure state is 0.  The following superficially valid syllogism is therefore fallacious:
\begin{enumerate}
\item If the electron is in the up state, $S = 0$.
\item The entropy of the electron is $S = \ln\,2$,
\item Therefore the electron is not in the up state.
\end{enumerate}
The correct conclusion is that the electron \emph{might} not be in the up state, a very different statement.  

Similarly, if the generalized entropy decreases somewhere on a null surface $N$, the proper conclusion to draw is not that $N$ is not a horizon, but that $N$ \emph{might not} be a horizon.  Let us take as a specific example the no-traversable-wormholes result from section \ref{black}.  A sufficiently adventurous spacefarer might not be deterred from attempted to crossing a wormhole simply because it is uncertain whether or not he will make it.  Suppose then that our intrepid hero lives in an asymptotically flat universe with state $\Psi$, and then jumps into the wormhole even though he only has a probability $1 > p > 0$ of successfully reaching the other side.  Assuming he \emph{does} reach the other side, it is then appropriate, at least on a forward-going basis, to project the state of the universe onto a new state $\Psi^\prime$ in which the wormhole jump certainly occurs, by using the projection operator $P$ onto the fact of the wormhole traversal:
\begin{equation}
| \Psi^\prime \rangle = \frac{P}{\sqrt{p}} | \Psi \rangle 
= \sqrt{p} | \Psi \rangle + \sqrt{1 - p} | \chi \rangle,
\end{equation}
where $\chi$ is some orthogonal state.

Here it is necessary to be careful.  As usual in quantum mechanical measurement, the state $\Psi^\prime$ will not be a good description of the state of the universe prior to the time of measurement.  In fact, since the additional branch $\chi$ of the superposition is defined using a future boundary condition, by the arguments in section \ref{OSL}, one expects the coarse-grained ordinary entropy of $\chi$ to be decreasing with time prior to the act of measurement.  However, it is not necessary to insist on $\Psi^\prime$ being the true state of the universe, or on wavefunction collapse being the correct interpretation of quantum mechanical measurements.  It is only required that $\Psi^\prime$ be a well-defined state in the theory, to which the fine-grained GSL must therefore apply.  One expects that $\Psi^\prime$ will be asymptotically flat since this boundary condition should not be affected by anything which goes on in the interior of spacetime.  Then 
$\Psi^\prime$ is an asymptotically flat spacetime in which there exists a traversable wormhole with probability 1, contradicting the result in section \ref{black}.  Similar arguments apply to the other trapped-surface no-go results in section \ref{app}.\footnote{In contrast, entropy fluctuations are relevant when applying the GSL to inflationary cosmology.  In an inflationary scenario, there is a scalar field $\Phi$ with some potential $V(\Phi)$, which gives rise to vacuum energy and a de-Sitter-like exponential expansion.  As this scalar field rolls down the potential, inflation comes to a halt.  The argument of eternal inflation is that quantum fluctuations sometimes push the field back up the potential and thus increase rather than decrease the vacuum energy \cite{eternal}.  Since an upward fluctuation results in a decrease in the horizon area, na\"{i}vely it would seem that the GSL forbids this process as well.  If so, the vacuum energy would be nonincreasing as time passes.  And if it cannot increase, one would generically expect it to decrease, and eventually exit inflation everywhere.

Assessing the validity of this argument requires a careful consideration of vacuum fluctuations.  The argument above that entropy fluctuations do not matter applies to asymptotically flat spacetimes.  De sitter space is different for two reasons: a) In de Sitter space there is a maximum value of the generalized entropy, so the necessary downward fluctuation $\Delta S$ of the horizon entropy is finite rather than infinite. b) In order to have eternal inflation, it is not necessary that any pre-selected region of spacetime remain inflating, but only that there exists some region that continues to do so.  If, over a given time interval, a Hubble volume increases in volume by a factor of $N$, it is only necessary to have a probability of about $1/N$ that inflation continue in each Hubble volume in order to keep inflation going somewhere.  If $\ln N > \Delta S$, then one would expect the necessary entropy fluctuation to occur in one of the $N$ regions.  It would be interesting to check whether this condition places significant constraints on eternal inflation scenarios.\label{inflation}}

It therefore follows from the GSL that the probability of forming a baby universe, a traversable wormhole, or restarting inflation in asymptotically flat or AdS spacetime, is \emph{exactly} zero.  This is a little surprising because one might have thought that these things could occur through quantum tunneling \cite{bubble, monopole}.  However, it is in accordance with the observation in Ref. \cite{myers} that one cannot restart inflation in AdS spacetimes in the context of AdS/CFT.

\subsection{Quantum Geometries}\label{quantgeo}

Since singularities are regions where quantum gravity effects might become important, one critical question is whether the GSL (as defined in section \ref{GSL}) continues to remain well-defined and true in the quantum gravity regime.  If not, then the quantum singularity theorem might break down just when it is needed, although many of the other applications of the Penrose theorem would continue to be useful.  Consider the following hierarchy of increasingly ``quantum'' treatments of spacetime:
\begin{enumerate}
\item Weak semiclassical perturbations.  This regime is simply quantum field theory in curved spacetime, plus infinitesimal corrections due to the gravitational deformation from matter.  (This regime justifies the `semiclassical expansion' used in Theorems 1,3 \&4 of section \ref{theorems}.)

\item Strong semiclassical effects.  In this regime one permits the quantum fields to have large gravitational effects, but neglects any quantum fluctuations in the metric, so that spacetime is still described by a smooth Lorentzian manifold.  This regime can be justified in some cases when there are a large number $N$ of matter fields and one takes $\hbar \to 0$ while holding $N\hbar$ fixed.  This suppresses graviton loops relative to matter loops.

\item Quantum Lorentzian manifold.  In this regime one takes into account the fluctuations in the metric, treating the metric as a quantum field living on a fixed, continuous topological space.  Some of the geometrical quantities now fail to commute.  This regime includes spacetimes which can be described by perturbative quantum gravity.  It is even possible, if the asymptotic safety scenario is true, that this regime might encompass a complete theory of quantum gravity \cite{safety}.

\item Something New.  Examples include causal sets \cite{causal}, the discrete spacetimes of loop quantum gravity \cite{lqg} or matrix theory \cite{matrix}, etc.  Here one cannot say anything nonspeculative except on the assumption of a specific model.
\end{enumerate}

\noindent Which of these regimes is physically relevant for the no-go results?

Not all of the results in this article require probing the quantum gravity regime.  For example, the absence of traversable wormholes (section \ref{black}) or warp drives (section \ref{warp}) is interesting even perturbatively, in the weak semiclassical regime (\#1).  That is because there exist classical solutions that are right on the edge of violating these results, so that a violation could be seen even perturbatively.  (For warp drives, consider pertrubations to the vacuum solution.  For traversable wormholes, perturb the eternal black hole.)\footnote{Of course, if these results continue to hold even in even more quantum spacetimes (regimes \#2-4), for the reasons suggested below, so much the better.}

But the usefulness of the no-go results concerning singularity theorems and baby universes (sections \ref{black}-\ref{begin}) depends on their applicability to the region near the singularity.  Consider a spacetime whose classical evolution has a singularity, and suppose that some resolution of the singularity \emph{were} possible.  Necessarily, any resolution of the singularity would have to involve nonperturbative effects, and therefore the weak semiclassical regime (\#1) will be insufficient to resolve the singularity.\footnote{As a general mathematical fact, when singularities in some function are resolved, the resolution tends to be nonperturbative.  At any finite order in perturbation theory the singularity typically gets more divergent rather than less divergent.  A simple example: if you Taylor expand the function $f(x) = 1/(x^2 + a^2)$ in $a$ around $a = 0$, each term in the Taylor series is progressively \emph{more} divergent with respect to $x$, but at finite positive values of $a$, there is no singularity.  I would like to thank Ed Witten for pointing out this issue.}

However, there is still hope for a useful result. The key thing to notice is that these results involve two distinct locations on the same null-surface $N$, separated by some null interval $\Delta \lambda$.  The region $X$ near the quantum trapped surface may be weakly coupled, even while the classically singular region $Y$  is strongly coupled.  Near $X$, it can be shown that the generalized entropy is decreasing somewhere on the null surface.  Then the GSL implies that $N$ cannot be a causal horizon.  Near $Y$, we use the fact that $N$ is not a horizon to show that there can be no infinite worldline $W$ in $Y$.

For example, in the case of the Big Bang result, the region $Y$ is the region near\footnote{and before, if the singularity is resolved} the classical would-be initial singularity when the universe was very hot and dense, while $X$ may be taken to be the present-day cosmology, which is very well-described by classical general relativity.  The analysis of region $X$ by itself can therefore be carried out entirely in the weak semiclassical regime (\#1).  Thus, the applications of Theorems 1 and 2 to region $X$ are insensitive to quantum gravity effects.  $X$ is also the only location at which we used a formula for the generalized entropy $S_\mathrm{gen}$, so quantum gravity corrections to the formula for $S_\mathrm{gen}$ are irrelevant to the validity of the result.

This leaves the analysis of the strongly coupled region $Y$, which might be in any of regimes $\#2-4$.  In order for the GSL to be well defined, it is necessary that the concept of a causal horizon still exist.  The notion is clearly defined in the strong semiclassical regime (\#2) due to the existence of a Lorentzian spacetime.  For a quantum Lorentzian manifold (\#3), the causal structure becomes fuzzy and thus one might worry about whether the causal horizon is defined.  Let us assume the following correspondence principles: A) that the requirement in general relativity that coordinates be smooth is merely for technical convenience, and that therefore one can consistently formulate general relativity to be covariant under the choice of arbitrary continuous coordinates, not just diffeomorphisms, B) that for any way of consistently gauge-fixing classical general relativity, there is a corresponding way to gauge-fix a quantum Lorentzian manifold, without introducing an anomaly into the true diffeomorphism-invariance of the theory.

Let there be some locus of points $L$ defined by some generally covariant prescription.  Then $\partial I^-(L)$ is a continuous (but not necessarily smooth) surface of codimension 1.  By assumption A above it is consistent to gauge-fix general relativity using a coordinate system in which one coordinate $x$ satisfies $x = 0$ at $\partial I^-(L)$.  In this coordinate system the horizon location is in a well-defined, fixed position in space, and therefore does not fluctuate in its position.  If $L$ is taken to be a future-infinite worldline defined by any coordinate invariant prescription, this shows that the notion of a future horizon is well-defined.  The GSL can then be defined to require that any such future causal horizon have nondecreasing entropy, at least in the semiclassical region $X$ where we know how to define $S_\mathrm{gen}$.

Even if the spacetime geometry is described by some new discrete structure (\#4), it still seems reasonable to believe that the notion of a future horizon may be well-defined, \emph{if} this discrete structure has a fundamental notion of causality built into it.  One can think of the location of a causal horizon as being defined by the way in which it divides spacetime points into exterior and interior regions.  The exterior of a future horizon can be defined as $J^-(W_\mathrm{fut})$, the causal past of a future-infinite worldline.  A worldline $W$ can be defined as a chain of points in causal sequence.  The only part of the definition which depends on anything other than a causal structure, is the requirement that $W_\mathrm{fut}$ be infinite.  But this can naturally be defined in a discrete geometry by requiring $W$ to consist of an infinite number of points.

So if the quantum geometry of the universe is a discrete causal set or anything richer, the notion of a future horizon should be well-defined, and it should be possible to ask whether the GSL is true.  Of course, it might turn out to be false.  For example, if quantum gravity violates Lorentz invariance, then the arguments of Ref. \cite{2speed} suggest that the GSL will be invalid.

However, it seems more elegant for the GSL to be true in quantum gravity.  This would explain the success of horizon thermodynamics in semiclassical general relativity.  Not only that, but by the results in sections \ref{warp} and \ref{time}, it would also ensure that the theory has positive energies and good causality.

\paragraph{Global Hyperbolicity.}

About half of the results in section \ref{app} assume global hyperbolicity, either directly or through the use of Theorems 2 or 4.  This includes the generalization of the Penrose singularity theorem (sections \ref{black} and \ref{begin}), some of the discussion about the thermodynamic beginning of the universe (\ref{begin}), and the prohibition of warp drives, and negative mass objects (\ref{warp}).  On the other hand, the prohibitions on viable baby universes, traversable wormholes, and restarting inflation (\ref{black}) use only the GSL, while the no-time machines result (\ref{time}) uses a weakened form of global hyperbolicity.

If the theory of quantum gravity is fully predictive, one expects some analogue of global hyperbolicity to be true, but it may not have quite the same implications as in general relativity.  Just because topology change is forbidden for continuous globally hyperbolic manifolds \cite{geroch}, does not necessarily mean it could not occur in discrete spacetimes.  Thus it is necessary to examine whether one expects theorems 2 and 4 to continue to hold.  Theorem 2 only depends on global hyperbolicity insofar as this is necessary to identify causal subsystems; it is therefore likely to hold in any theory with causality.

Theorem 4 depends on global hyperbolicity in a more subtle way.  The basic causality assumption underlying Theorem 4 is that an outward moving causal surface on a noncompact spatial slice cannot come to an end without encountering a boundary of the spacetime.  This might happen in two different ways: (a) A noncompact space could become compact as a result of time evolution, or (b) A noncompact space could split into two regions, one of them compact, and the other noncompact.  It seems unlikely that quantum topology change could permit (a), since it would require an ``infinitely large'' tunneling event.  Scenario (b) is the disconnected baby universe scenario, which is forbidden by the GSL \emph{without} using global hyperbolicity.\footnote{Additionally, (b) raises potential problems with causality. If one runs the process in time-reverse, one finds that two completely unrelated regions of spacetime spontaneously join together.  This seems to be an extreme violation of locality.  But see Ref. \cite{coleman} for a possible way around this argument.}  Accepting these arguments against (a) and (b), it is not unlikely that an analogue of Theorem 4 may well apply in full quantum gravity.

Thus there is a reasonable possibility that the Penrose singularity theorem can be proven even in the context of full quantum gravity.  This would go against the conventional wisdom that the singularities are an symptom of the incompleteness of the classical theory, and are resolved quantum mechanically.  However, it should be pointed out that just because there are singularities in the sense that spacetime comes to an end in some places, does not mean that there are any physical quantities which become infinite at the singularity.  A discrete geometry might still resolve the singularity in the latter sense by cutting off the spacetime at distances shorter than the Planck scale.

\section{Conclusion}

It has been shown above that, under the assumption that spacetime is a globally hyperbolic Lorentzian manifold, the fine-grained GSL requires black holes and infinite FRW universes to have singularities, and places severe constraints on baby universes and any cosmology prior to the Big Bang.  It additionally prevents asymptotically flat spacetimes from having negative ADM masses, warp drives or traversable wormholes, or developing time machines or inflating regions.  In all of these cases, theorems of classical general relativity have been extended to semiclassical settings by using the GSL as a premise instead of the null energy condition.  The notion of a ``trapped surface'' still persists in this quantum setting, and ensures that these qualitative features of semiclassical gravity are the same as those of classical general relativity.  (It should be reiterated, however, that the GSL has only been proven in limited regimes \cite{10proofs}, and that there might be other reasonable ways to formulate the GSL besides the one given in section \ref{GSL}.)

There are also some---necessarily speculative---indications that these results might hold in a full theory of quantum gravity.  Although the semiclassical approximation was used to derive some of the no-go results, it was only used in nearly classical regions, either a large distance or a long time away from high curvature quantum gravity regions.  Other than the GSL itself, the only assumptions made about the high curvature region were that spacetime continues to have some of the same primitive properties as a Lorentzian spacetime: a notion of causality used to define horizons, a notion of predictivity analogous to global hyperbolicity, distinctions between finite and infinite lengths, and compact and noncompact regions.  Given the successes of horizon thermodynamics, it is natural to suppose that the GSL holds even at the level of quantum gravity, and thus that not all singularities are resolved in quantum gravity.

The statistical mechanical argument for a beginning in time, based on the fact that entropy decreases when going to the past, was also generalized to an argument from the coarse-grained GSL.  Together with the singularity theorem, this leads to a prima facie argument that time had a true beginning at the Big Bang some 13.7 billion years ago.  In section \ref{begin}, in order to make a plausible GSL-satisfying cosmology with an infinite past, it was necessary to postulate both that the cosmos is spatially finite, and that the arrow of time was reversed before some time $t_0$.  This kind of bounce evades both the singularity and thermodynamic arrow constraints, but still has in some sense a thermodynamic `beginning' in time at the moment of lowest entropy.  That is, both the past and the future would be explained in terms of the low entropy state at $t_0$, while the state at $t_0$ would itself have no explanation in terms of anything to the future or the past.  (Thus the moment $t_0$ would seem to raise the same sorts of philosophical questions that any other sort of beginning in time would.)

The fact that the no-go results forbid various processes with probability 0 is interesting because it goes against the usual experience in quantum field theory that anything not forbidden by kinematics or conservation laws must occur with some nonzero probability.  This suggests that there may be a formulation of quantum geometry based on horizon thermodynamics in which these constraints seem more natural.

The notion of a quantum trapped surface from Theorem 4 may be a clue here.  If we think that horizon thermodynamics works because of the statistical mechanics of the quantum gravity degrees of freedom near or on the horizon, what should we make of the fact that on certain surfaces, the entropy does decrease?  It is as though each null surface must either choose to be a causal horizon and behave in certain respects like a closed system, or else violate the second law and be punished for it by coming to an end in a finite time.  Can this basic dichotomy be explained somehow from the perspective of the microscopic horizon degrees of freedom?

\section*{Acknowledgements}
\small
This work was supported by NSF grants PHY-0601800 and PHY-0903572, the Maryland Center for Fundamental Physics, the Perimeter Institute for Theoretical Physics, and the Simons Foundation.  I am grateful for conversations with Ted Jacobson, William Donnelly, Sudipta Sarkar, Latham Boyle, Rob Myers, Don Marolf, Robert Brandenberger, Netta Engelhardt, Ed Witten, and for detailed comments from anonymous referees.
\normalsize

\section*{Appendix}

This appendix will prove a theorem used in section \ref{OSL} to prove that entropy increases in the context of ordinary thermodynamics.  There, the approach was to model our uncertainty about time evolution using a mixture of possible unitary operators, acting on any separable Hilbert Space.

Let $dU$ be a probability distribution over the space of possible unitary operators, such that the total probability is $1$:
\begin{equation}
\int_U dU = 1.
\end{equation}
Then the operation to be performed on the state $\rho$ is
\begin{equation}
\rho \to T(\rho) \equiv \int_U U \rho\,U^\dagger\,dU.
\end{equation}
$T$ is a linear map from the space of all density matrices to itself.  There are some consequences of the fact that the map is just a sum over unitary operators:  a) By conservation of probability, $T$ must preserve the trace of $\rho$.  b) Since probabilities cannot be negative, $T$ must map states with nonnegative eigenvalues to other states with nonnegative eigenvalues.  Furthermore, c) Since every $U$ preserves the identity state $I$, $T(I) = I$.  

Such maps cannot decrease the entropy, a fact which has been derived from the Uhlmann theory of mixing \cite{Uhlmann}, by way of an even stronger statement about probability eigenvalues: namely that there is no way of transferring probability from a smaller eigenvalue to a larger one.  

More precisely, for all natural numbers $i$, the sum of the $i$th largest eigenvalues of the final density matrix must be no greater than the $i$th largest eigenvalues of the initial density matrix.  In other words, if we consider the spectrum of probability eigenvalues of $\rho$, a probability eigenvalue can only increase in value if it does so at the expense of eigenvalues with greater probability.  This stronger statement is more powerful than simply asserting that entropy increases, because it yields a separate statement for each number $i$.

One might worry about how to interpret ``the sum of the ith largest eigenvectors'' if $\rho$ is degenerate.  There is then an ambiguity as to a) which basis to use to count eigenvectors, and b) how to order the eigenvalues seeing as some of them are equal.   However, because by definition all of the degenerate eigenvalues have the same value, it makes no difference to the sum how we partition the degenerate states so long as we make some choice.

A proof of this result follows:
\newline \textbf{Definitions:}

$\rho$ is the density matrix,

$T()$ is the trace and identity preserving, positive linear map, 
 
$I$ is the identity matrix, 

$M = \rho - p_{i} I$, where $p_{i}$ is the ith largest eigenvalue, 

$P$ projects onto the $i$th largest eigenvalues of $\rho$, and 

$Q$ projects onto the $i$th largest eigenvalues of $T(\rho)$.\newline
\textbf{Theorem 5:}  The theorem to be proven can now be stated as follows:
\begin{equation}
\mathrm{tr}(Q\:T(\rho)\:Q) \le \mathrm{tr}(P \rho \, P)
\end{equation}
\textbf{Proof:} Since $P$ commutes with $\rho$ and thus $M$, we may write
\begin{equation}
M = PMP + (1-P)M(1-P) = A - B
\end{equation}
where $A$ and $B$ are manifestly positive.
Now by positivity,
\begin{equation}
\mathrm{tr}(Q\:T(A)\:Q) \le \mathrm{tr}(T(A)) = \mathrm{tr}(A)
\end{equation}
since a partial trace of a positive matrix cannot give more than  
the full trace, while
\begin{equation}
\mathrm{tr}(Q\:T(B)\:Q) \ge 0
\end{equation}
because each operation preserves positivity.
Therefore by linearity, 
\begin{equation}
\mathrm{tr}(Q\:T(M)\:Q) \le \mathrm{tr}(A) = \mathrm{tr}(PMP) \label{finalstep}
\end{equation}
which shows that the sum of the $i$ larger eigenvalues can only  
decrease.  But $M$ and $\rho$ only differ by $I = T(I)$ so the result  
holds for $\rho$ as well.  Q.E.D.

\textbf{Corollary 5.1:} Any quantity expressible as $\mathrm{tr}(f(\rho))$, where $f$ is any convex function, is nondecreasing.  This is implied by the fact that the probability eigenvalues can only evolve towards equalization \cite{Uhlmann}.

This stronger statement about probability eigenvalues is equivalent to saying that any convex function of the probability eigenvalues is nondecreasing.  Pick the convex function:
\begin{eqnarray}
         f(p) = 0 \phantom{p_i - p}          & p \le p_i \label{1st} \\ 
              = p_i - p \phantom{0}          & p \ge p_i \label{2nd}
\end{eqnarray}
This function must satisfy
\begin{equation}
\mathrm{tr} f(T(\rho)) \ge \mathrm{tr} f(\rho)
\end{equation}
which implies that the probability eigenvalues can only equalize.

\end{document}